\documentclass[12pt,aps,showpacs,preprintnumbers,superscriptaddress,nofootinbib]
{revtex4}
\usepackage{graphicx}

\begin{document}

%%%%%%%%%%%%%%%%%%%%%%%%%%%%%%%%%%%%%%%%%%%%%

\newcommand{\sst}[1]{{\scriptscriptstyle #1}}
\newcommand{\beq}{\begin{equation}}
\newcommand{\eeq}{\end{equation}}
\newcommand{\beqa}{\begin{eqnarray}}
\newcommand{\eeqa}{\end{eqnarray}}
\newcommand{\dida}[1]{/ \!\!\! #1}
\renewcommand{\Im}{\mbox{\sl{Im}}}
\renewcommand{\Re}{\mbox{\sl{Re}}}
\def\simge{\hspace*{0.2em}\raisebox{0.5ex}{$>$}
     \hspace{-0.8em}\raisebox{-0.3em}{$\sim$}\hspace*{0.2em}}
\def\simle{\hspace*{0.2em}\raisebox{0.5ex}{$<$}
     \hspace{-0.8em}\raisebox{-0.3em}{$\sim$}\hspace*{0.2em}}
\def\dn{{d_n}}
\def\de{{d_e}}
\def\datom{{d_{\sst{A}}}}
\def\grhobar{{{\bar g}_\rho}}
\def\gpibar{{{\bar g}_\pi^{(I) \prime}}}
\def\gpibarz{{{\bar g}_\pi^{(0) \prime}}}
\def\gpibaro{{{\bar g}_\pi^{(1) \prime}}}
\def\gpibart{{{\bar g}_\pi^{(2) \prime}}}
\def\mn{{m_{\sst{N}}}}
\def\mx{{M_X}}
\def\mrho{{m_\rho}}
\def\qpv{{Q_{\sst{W}}}}
\def\lamtv{{\Lambda_{\sst{TVPC}}}}
\def\lamtvs{{\Lambda_{\sst{TVPC}}^2}}
\def\lamtvc{{\Lambda_{\sst{TVPC}}^3}}

%     \hspace{-0.8em}\raisebox{-0.3em}{$\sim$}\hspace*{0.2em}}
\def\bra#1{{\langle#1\vert}}
\def\ket#1{{\vert#1\rangle}}
\def\coeff#1#2{{\scriptstyle{#1\over #2}}}
\def\undertext#1{{$\underline{\hbox{#1}}$}}
\def\hcal#1{{\hbox{\cal #1}}}
\def\sst#1{{\scriptscriptstyle #1}}
\def\eexp#1{{\hbox{e}^{#1}}}
\def\rbra#1{{\langle #1 \vert\!\vert}}
\def\rket#1{{\vert\!\vert #1\rangle}}

\def\lsim{{ <\atop\sim}}
\def\gsim{{ >\atop\sim}}
\def\nubar{{\bar\nu}}
\def\psibar{{\bar\psi}}
\def\Gmu{{G_\mu}}
\def\alr{{A_\sst{LR}}}
\def\wpv{{W^\sst{PV}}}
\def\evec{{\vec e}}
\def\notq{{\not\! q}}
\def\notl{{\not\! \ell}}
\def\notk{{\not\! k}}
\def\notp{{\not\! p}}
\def\notpp{{\not\! p'}}
\def\notder{{\not\! \partial}}
\def\notcder{{\not\!\! D}}
\def\notA{{\not\!\! A}}
\def\notv{{\not\!\! v}}
\def\Jem{{J_\mu^{em}}}
\def\Jana{{J_{\mu 5}^{anapole}}}
\def\nue{{\nu_e}}
\def\mn{{m_{\sst{N}}}}
\def\mns{{m^2_{\sst{N}}}}
\def\me{{m_e}}
\def\mes{{m^2_e}}
\def\mq{{m_q}}
\def\mqs{{m_q^2}}
\def\mw{{M_{\sst{W}}}}
\def\mz{{M_{\sst{Z}}}}
\def\mzs{{M^2_{\sst{Z}}}}
\def\ubar{{\bar u}}
\def\dbar{{\bar d}}
\def\sbar{{\bar s}}
\def\qbar{{\bar q}}
\def\sstw{{\sin^2\theta_{\sst{W}}}}
\def\gv{{g_{\sst{V}}}}
\def\ga{{g_{\sst{A}}}}
\def\pv{{\vec p}}
\def\pvs{{{\vec p}^{\>2}}}
\def\ppv{{{\vec p}^{\>\prime}}}
\def\ppvs{{{\vec p}^{\>\prime\>2}}}
\def\qv{{\vec q}}
\def\qvs{{{\vec q}^{\>2}}}
\def\xv{{\vec x}}
\def\xpv{{{\vec x}^{\>\prime}}}
\def\yv{{\vec y}}
\def\tauv{{\vec\tau}}
\def\sigv{{\vec\sigma}}

\def\sst#1{{\scriptscriptstyle #1}}
\def\gpnn{{g_{\sst{NN}\pi}}}
\def\grnn{{g_{\sst{NN}\rho}}}
\def\gnnm{{g_{\sst{NNM}}}}
\def\hnnm{{h_{\sst{NNM}}}}
\def\xivz{{\xi_\sst{V}^{(0)}}}
\def\xivt{{\xi_\sst{V}^{(3)}}}
\def\xive{{\xi_\sst{V}^{(8)}}}
\def\xiaz{{\xi_\sst{A}^{(0)}}}
\def\xiat{{\xi_\sst{A}^{(3)}}}
\def\xiae{{\xi_\sst{A}^{(8)}}}
\def\xivtez{{\xi_\sst{V}^{T=0}}}
\def\xivteo{{\xi_\sst{V}^{T=1}}}
\def\xiatez{{\xi_\sst{A}^{T=0}}}
\def\xiateo{{\xi_\sst{A}^{T=1}}}
\def\xiva{{\xi_\sst{V,A}}}
\def\rvz{{R_{\sst{V}}^{(0)}}}
\def\rvt{{R_{\sst{V}}^{(3)}}}
\def\rve{{R_{\sst{V}}^{(8)}}}
\def\raz{{R_{\sst{A}}^{(0)}}}
\def\rat{{R_{\sst{A}}^{(3)}}}
\def\rae{{R_{\sst{A}}^{(8)}}}
\def\rvtez{{R_{\sst{V}}^{T=0}}}
\def\rvteo{{R_{\sst{V}}^{T=1}}}
\def\ratez{{R_{\sst{A}}^{T=0}}}
\def\rateo{{R_{\sst{A}}^{T=1}}}
\def\mro{{m_\rho}}
\def\mks{{m_{\sst{K}}^2}}
\def\mpi{{m_\pi}}
\def\mpis{{m_\pi^2}}
\def\mom{{m_\omega}}
\def\mphi{{m_\phi}}
\def\Qhat{{\hat Q}}
\def\FOS{{F_1^{(s)}}}
\def\FTS{{F_2^{(s)}}}
\def\GAS{{G_{\sst{A}}^{(s)}}}
\def\GES{{G_{\sst{E}}^{(s)}}}
\def\GMS{{G_{\sst{M}}^{(s)}}}
\def\GATEZ{{G_{\sst{A}}^{\sst{T}=0}}}
\def\GATEO{{G_{\sst{A}}^{\sst{T}=1}}}
\def\mdax{{M_{\sst{A}}}}
\def\mustr{{\mu_s}}
\def\rsstr{{r^2_s}}
\def\rhostr{{\rho_s}}
\def\GEG{{G_{\sst{E}}^\gamma}}
\def\GEZ{{G_{\sst{E}}^\sst{Z}}}
\def\GMG{{G_{\sst{M}}^\gamma}}
\def\GMZ{{G_{\sst{M}}^\sst{Z}}}
\def\GEn{{G_{\sst{E}}^n}}
\def\GEp{{G_{\sst{E}}^p}}
\def\GMn{{G_{\sst{M}}^n}}
\def\GMp{{G_{\sst{M}}^p}}
\def\GAp{{G_{\sst{A}}^p}}
\def\GAn{{G_{\sst{A}}^n}}
\def\GA{{G_{\sst{A}}}}
\def\GETEZ{{G_{\sst{E}}^{\sst{T}=0}}}
\def\GETEO{{G_{\sst{E}}^{\sst{T}=1}}}
\def\GMTEZ{{G_{\sst{M}}^{\sst{T}=0}}}
\def\GMTEO{{G_{\sst{M}}^{\sst{T}=1}}}
\def\lamd{{\lambda_{\sst{D}}^\sst{V}}}
\def\lamn{{\lambda_n}}
\def\lams{{\lambda_{\sst{E}}^{(s)}}}
\def\bvz{{\beta_{\sst{V}}^0}}
\def\bvo{{\beta_{\sst{V}}^1}}
\def\Gdip{{G_{\sst{D}}^\sst{V}}}
\def\GdipA{{G_{\sst{D}}^\sst{A}}}
\def\fks{{F_{\sst{K}}^{(s)}}}
\def\FIS{{F_i^{(s)}}}
\def\fpi{{F_\pi}}
\def\fk{{F_{\sst{K}}}}
\def\RAp{{R_{\sst{A}}^p}}
\def\RAn{{R_{\sst{A}}^n}}
\def\RVp{{R_{\sst{V}}^p}}
\def\RVn{{R_{\sst{V}}^n}}
\def\rva{{R_{\sst{V,A}}}}
\def\xbb{{x_B}}
\def\mlq{{M_{\sst{LQ}}}}
\def\mlqs{{M_{\sst{LQ}}^2}}
\def\lscal{{\lambda_{\sst{S}}}}
\def\lvect{{\lambda_{\sst{V}}}}
\def\PR#1{{{\em   Phys. Rev.} {\bf #1} }}
\def\PRC#1{{{\em   Phys. Rev.} {\bf C#1} }}
\def\PRD#1{{{\em   Phys. Rev.} {\bf D#1} }}
\def\PRL#1{{{\em   Phys. Rev. Lett.} {\bf #1} }}
\def\NPA#1{{{\em   Nucl. Phys.} {\bf A#1} }}
\def\NPB#1{{{\em   Nucl. Phys.} {\bf B#1} }}
\def\AoP#1{{{\em   Ann. of Phys.} {\bf #1} }}
\def\PRp#1{{{\em   Phys. Reports} {\bf #1} }}
\def\PLB#1{{{\em   Phys. Lett.} {\bf B#1} }}
\def\ZPA#1{{{\em   Z. f\"ur Phys.} {\bf A#1} }}
\def\ZPC#1{{{\em   Z. f\"ur Phys.} {\bf C#1} }}
\def\etal{{{\em   et al.}}}
\def\delalr{{{delta\alr\over\alr}}}
\def\pbar{{\bar{p}}}
\def\lamchi{{\Lambda_\chi}}
\def\qw0{{Q_{\sst{W}}^0}}
\def\qwp{{Q_{\sst{W}}^P}}
\def\qwn{{Q_{\sst{W}}^N}}
\def\qwe{{Q_{\sst{W}}^e}}
\def\qem{{Q_{\sst{EM}}}}
\def\gae{{g_{\sst{A}}^e}}
\def\gve{{g_{\sst{V}}^e}}
\def\gvf{{g_{\sst{V}}^f}}
\def\gaf{{g_{\sst{A}}^f}}
\def\gvu{{g_{\sst{V}}^u}}
\def\gau{{g_{\sst{A}}^u}}
\def\gvd{{g_{\sst{V}}^d}}
\def\gad{{g_{\sst{A}}^d}}
\def\gvftil{{\tilde g_{\sst{V}}^f}}
\def\gaftil{{\tilde g_{\sst{A}}^f}}
\def\gvetil{{\tilde g_{\sst{V}}^e}}
\def\gaetil{{\tilde g_{\sst{A}}^e}}
\def\gvqtil{{\tilde g_{\sst{V}}^e}}
\def\gaqtil{{\tilde g_{\sst{A}}^e}}
\def\gvutil{{\tilde g_{\sst{V}}^e}}
\def\gautil{{\tilde g_{\sst{A}}^e}}
\def\gvdtil{{\tilde g_{\sst{V}}^e}}
\def\gadtil{{\tilde g_{\sst{A}}^e}}
\def\delp{{\delta_P}}
\def\delzp{{\delta_{00}}}
\def\deld{{\delta_\Delta}}
\def\dele{{\delta_e}}
\def\lnew{{{\cal L}_{\sst{NEW}}}}
\def\osffp{{{\cal O}_{7a}^{ff'}}}
\def\oszg{{{\cal O}_{7c}^{Z\gamma}}}
\def\osgg{{{\cal O}_{7b}^{g\gamma}}}

%%%%%%%%%%%%%%%%%%%%%%%%%%%%%%%%%%%%%%%%%%%%%

\def\slash#1{#1\!\!\!{/}}
\def\beq{\begin{eqnarray}}
\def\eeq{\end{eqnarray}}
\def\bea{\begin{eqnarray*}}
\def\eea{\end{eqnarray*}}
\def\NCA{\em Nuovo~Cimento}
\def\IJMP{\em Intl.~J.~Mod.~Phys.}
\def\NP{\em Nucl.~Phys.}
\def\PLB{{\em Phys.~Lett.}~B}
\def\JETPLett{{\em JETP Lett.}}
\def\PRL{\em Phys.~Rev.~Lett.}
\def\MPL{\em Mod.~Phys.~Lett.}
\def\PRD{{\em Phys.~Rev.}~D}
\def\PR{\em Phys.~Rev.}
\def\PRP{\em Phys.~Rep.}
\def\ZPC{{\em Z.~Phys.}~C}
\def\PTP{{\em Prog.~Theor.~Phys.}}
% Some other macros used in the sample text
\def\Baryon{{\rm B}}
\def\Lepton{{\rm L}}
\def\sbar{\overline}
\def\stilde{\widetilde}
\def\st{\scriptstyle}
\def\sst{\scriptscriptstyle}
\def\vac{|0\rangle}
\def\argh{{{\rm arg}}}
\def\G{\stilde G}
\def\Wmess{W_{\rm mess}}
\def\NI{\stilde N_1}
\def\antivac{\langle 0|}
\def\infinity{\infty}
\def\mco{\multicolumn}
\def\epp{\epsilon^{\prime}}
\def\psibar{\overline\psi}
\def\nmess{N_5}
\def\chibar{\overline\chi}
\def\lagr{{\cal L}}
\def\drbar{\overline{\rm DR}}
\def\msbar{\overline{\rm MS}}
\def\conj{{{\rm c.c.}}}
\def\Et{{\slashchar{E}_T}}
\def\Etot{{\slashchar{E}}}
\def\mZ{m_Z}
\def\MPlanck{M_{\rm P}}
\def\mW{m_W}
\def\cbeta{c_{\beta}}
\def\sbeta{s_{\beta}}
\def\cW{c_{W}}
\def\sW{s_{W}}
\def\deltaeps{\delta}
\def\sigmabar{\overline\sigma}
\def\epsilonbar{\overline\epsilon}
\def\vep{\varepsilon}
\def\ra{\rightarrow}
\def\half{{1\over 2}}
\def\ko{K^0}
\def\be{\beq}
\def\ee{\eeq}
\def\bea{\begin{eqnarray}}
\def\eea{\end{eqnarray}}
\def\alr{A_{\sst{LR}}}

%  \gsim and \lsim provide >= and <= signs.
\def\centeron#1#2{{\setbox0=\hbox{#1}\setbox1=\hbox{#2}\ifdim
\wd1>\wd0\kern.5\wd1\kern-.5\wd0\fi
\copy0\kern-.5\wd0\kern-.5\wd1\copy1\ifdim\wd0>\wd1
\kern.5\wd0\kern-.5\wd1\fi}}
\def\ltap{\;\centeron{\raise.35ex\hbox{$<$}}{\lower.65ex\hbox{$\sim$}}\;}
\def\gtap{\;\centeron{\raise.35ex\hbox{$>$}}{\lower.65ex\hbox{$\sim$}}\;}
\def\gsim{\mathrel{\gtap}}
\def\lsim{\mathrel{\ltap}}

%%%%%%%%%%%%%%%%%%%%%%%%%%%%%%%%%%%%%%%
%  Slash character...
\def\slashchar#1{\setbox0=\hbox{$#1$}           % set a box for #1
   \dimen0=\wd0                                 % and get its size
   \setbox1=\hbox{/} \dimen1=\wd1               % get size of /
   \ifdim\dimen0>\dimen1                        % #1 is bigger
      \rlap{\hbox to \dimen0{\hfil/\hfil}}      % so center / in box
      #1                                        % and print #1
   \else                                        % / is bigger
      \rlap{\hbox to \dimen1{\hfil$#1$\hfil}}   % so center #1
      /                                         % and print /
   \fi}                                        %

%%EXAMPLE:  $\slashchar{E}$ or $\slashchar{E}_{t}$
\setcounter{tocdepth}{2}

%%%%%%%%%%%%%%%%%%%%%%%%%%%%%%%%%%%%%%%%%%%%%

\preprint{CALT-68-2430}
\preprint{MAP-289}
\preprint{hep-ph/0303026}

\title{Probing Supersymmetry with Parity-Violating Electron Scattering}

\author{A. Kurylov}
\affiliation{
California Institute of Technology,
Pasadena, CA 91125\ USA}

\author{M.J. Ramsey-Musolf}
\affiliation{
California Institute of Technology,
Pasadena, CA 91125\ USA}
\affiliation{
Department of Physics, University of Connecticut,
Storrs, CT 06269\ USA}
\affiliation{
Institute for Nuclear Theory, University of Washington,
Seattle, WA
98195\ USA
}

\author{S.~Su}
\affiliation{
California Institute of Technology,
Pasadena, CA 91125\ USA}

\begin{abstract}

We compute the one-loop supersymmetric (SUSY) contributions to the weak
charges of the electron ($Q_W^e$), proton ($Q_W^p$), and cesium nucleus
($Q_W^{\rm Cs}$) in the Minimal
Supersymmetric Standard Model (MSSM). Such contributions can generate
several percent
corrections to the corresponding Standard Model values. The magnitudes of the
SUSY loop corrections to $Q_W^e$ and $Q_W^p$ are
correlated over nearly all of the MSSM parameter space and result in an
increase in the magnitudes of these weak charges. In contrast, the effects on
$Q_W^{\rm Cs}$ are
considerably smaller and are equally likely to increase or decrease its
magnitude.
Allowing for R-parity violation can lead to opposite sign relative shifts
in $Q_W^e$ and $Q_W^p$, normalized to the corresponding
Standard Model values. A comparison of $Q_W^p$ and $Q_W^e$ measurements
could help distinguish between different SUSY scenarios.

\end{abstract}


\maketitle

\pagenumbering{arabic}

\section{Introduction}
\label{sec:intro}

The search for physics beyond the Standard Model (SM) of electroweak and
strong interactions is
a primary objective for particle and nuclear physics. Historically,
parity-violating (PV)
interactions have played an important role in elucidating the structure of
the electroweak
interaction. In the 1970's,  PV deep inelastic scattering (DIS) measurements
performed at the Stanford Linear Accelerator Center (SLAC) confirmed the SM
prediction for the structure of weak neutral current
interactions \cite{SLAC}. These results were
consistent with a value for the weak mixing angle given by
$\sin^2\theta_W\approx 1/4$, implying
a tiny $V$(electron)$\times A$(quark) neutral current interaction.
Subsequent PV measurements --
performed at both very low scales using atoms as well as at the $Z$-pole
in $e^+e^-$
annihilation -- have  been remarkably consistent with the results of the
SLAC DIS
measurement\cite{SLAC}.

More recently, the results of cesium atomic
parity-violation (APV) \cite{Ben99} and deep inelastic $\nu$- ($\bar\nu$-)
nucleus scattering\cite{NuTeV}  have been interpreted as determinations of the
scale-dependence of $\sin^2\theta_W$. The SM predicts how this quantity should
depend on the momentum transfer squared ($q^2$) of a given process
\footnote{
The weak mixing angle and its
$q^2$ evolution are renormalization scheme-dependent. Here, we use the
$\overline{\rm MS}$ scheme for the SM and the $\overline{\rm DR}$ scheme
for supersymmetric extensions of the SM.
}.
The cesium APV result appears to be
consistent with the SM prediction for $q^2\approx 0$, whereas the neutrino
DIS measurement implies a $+3\sigma$ deviation at $|q^2|\sim 20$
(GeV$/c)^2$. If
conventional hadron structure effects are ultimately unable to
account for the NuTeV \lq\lq anomaly", the results of this precision
measurement would point
to new physics.

In light of this situation, two new measurements involving polarized
electron scattering
have taken on added interest: PV M\"oller ($ee$) scattering at
SLAC\cite{slac} and elastic,  PV
$ep$ scattering at the Jefferson Lab (JLab)\cite{qweak}. In the absence of
new physics, both
measurements could be used to determine $\sin^2\theta_W$ at the same scale
[$|q^2|\approx
0.03$ $({\rm GeV}/c)^2$] -- falling between the scales relevant
to the APV and neutrino DIS measurements -- with comparable precision in
each case
\footnote{In practice, the PV $ep$ experiment will actually
provide a value for $\sin^2\theta_W(q^2=0)$, as discussed in Ref.
\cite{Erl-MJRM-Kur02}.
} ($\Delta\sin^2\theta_W\approx 7\times 10^{-4}$).
Any
significant deviation from the SM prediction for $\sin^2\theta_W$ at this
scale  would provide
striking evidence for new physics, particularly if
both measurements report a deviation. On the other hand, agreement would
imply that the
most likely explanation for the neutrino DIS result involves hadron
structure effects within the SM.

In this paper, we analyze the prospective implications of the
parity-violating electron scattering (PVES) measurements for supersymmetry
(SUSY).
Although no supersymmetric
particle has yet been discovered, there exists strong theoretical
motivation for
believing that SUSY is a
component of the \lq\lq new" Standard Model. For example, the existence of
low-energy SUSY is a prediction of many string theories; it offers a
solution to the
hierarchy problem; and it results in coupling unification
close to the Planck scale. In addition, if R-parity is conserved (see below),
SUSY provides an excellent candidate for cold dark matter, the lightest
neutralino (see Ref.~\cite{kamion-cosm} for a review). The existence of such dark
matter is required in most cosmological models.
In light of such arguments, it is clearly of interest to determine
what insight about SUSY--if any--the new PVES measurements might provide.

In the simplest version of SUSY -- the Minimal Supersymmetric Standard
Model (MSSM) with conserved R-parity \cite{haber-kane}
-- low-energy precision observables experience SUSY only via tiny loop
effects involving
virtual supersymmetric particles. The requirement of baryon minus lepton number ($B-L$)
conservation leads to conservation of the R-parity quantum number,
$P_R=(-1)^{2S+3(B-L)}$, where $S$ denotes spin. Every SM particle has
$P_R=+1$ while
the corresponding superpartner, whose spin differs by $1/2$ unit, has $P_R=-1$.
Conservation of $P_R$ implies that every vertex has an even number of
superpartners.
Consequently, for processes like $ee\to ee$ and $ep\to ep$, all
superpartners must live in
loops, which generate corrections -- relative to the SM amplitude -- of order
$(\alpha/\pi)(M/{\tilde M})^2\sim 10^{-3}$ (where $M$ denotes a SM particle
mass and
${\tilde M}$ is a
superpartner mass).  Generally speaking, then, low-energy experiments must
probe an observable
with a precision of
few tenths of a percent  or better in order to discern SUSY loop effects.
Low-energy
charged current experiments have already reached such levels of precision,
and the
corresponding implications of these experiments for the MSSM have been
discussed
elsewhere\cite{kurylov02}.

In the case of PV $ee$ and elastic $ep$ scattering, the precision needed to
probe SUSY loop effects is roughly an order of magnitude less stringent,
owing to a fortuitous suppression of the SM PV asymmetries, $\alr$. At
leading order in
$q^2$, the $A(e)\times V(f)$ contributions to $\alr$ are governed by
$Q_W^f$, the
\lq\lq weak charge" of the target fermion, $f$.
 The weak charge of a particle $f$ is defined as the strength
of the effective $A(e)\times V(f)$ interaction:
\begin{eqnarray}
\label{eq:Leff}
{\cal L}_{EFF}^{ef}=-{G_\mu\over 2\sqrt 2}Q_W^f {\bar e} \gamma_\mu\gamma_5
e {\bar f}\gamma_\mu f~.
\end{eqnarray}
At tree-level in the SM the weak charges of both the electron and the
proton are suppressed:
$Q_W^p=-Q_W^e=1-4\sin^2\theta_W\approx 0.1$. One-loop SM electroweak radiative
corrections further reduce this tiny number, leading to the predictions
$Q_W^e=-0.0449$\cite{Mar96,Erl-MJRM-Kur02} and
$Q_W^p=0.0716$\cite{Erl-MJRM-Kur02}. The  factor of $\gsim$10 suppression of these
couplings
in the SM renders them
more transparent to the possible effects of new physics. Consequently,
experimental precision of order a few percent, rather than a
few tenths of a percent, is needed to probe SUSY loop corrections.
(Theoretical uncertainties associated with QCD corrections to $Q_W^{e,p}$
are considerably smaller \cite{Mar96,Erl-MJRM-Kur02}).

In analyzing these SUSY loop contributions to $Q_W^{e,p}$, we carry out a
model-independent treatment, avoiding the choice of a
specific mechanism for SUSY-breaking mediation. While most analyses of
precision electroweak observables have been performed using one or more
widely-used models for SUSY-breaking mediation, the  generic features of
the superpartner spectrum implied by such models
may not be consistent with precision data \cite{kurylov02}. Consequently,
we wish to
determine the possible impact of SUSY on the two PVES measurements for all
phenomenologically acceptable choices of the MSSM parameters, even if such
choices lie outside the purview of standard SUSY-breaking models. In doing so,
we follow the spirit of Ref.~\cite{kur-rm-su-nutev}, where a similar
analysis of
SUSY loop effects in $\nu$ ($\bar \nu$)-nucleus scattering was performed.

In the case of PV electron scattering,
we find that the magnitude of SUSY loop effects could be as large as the
proposed experimental uncertainties for the $Q_W^{e}$ and $Q_W^p$
measurements (8\% and 4\%,
respectively \cite{slac,qweak}). Moreover,
the relative sign of the effect (compared to the SM prediction) in both
cases is correlated -- and positive -- over nearly all available SUSY
parameter space. To our knowledge, such correlation is specific to the MSSM
(with R-parity conservation),
making it a potential low-energy signature of this new physics scenario. We
also find that the SUSY
loop effects on $Q_W^{\rm Cs}$, the weak charge of the cesium atom measured
in APV, is much
less pronounced. Thus, the present agreement between the experimental value
for $Q_W^{\rm Cs}$
and the SM prediction does not preclude the presence of relatively large
effects in the PV electron
scattering asymmetries.

We also investigate a scenario where $P_R$ is not conserved.
We find that, in contrast to the $P_R$-conserving SUSY, the relative
sign of the effect (compared to the SM prediction) is always negative
for $Q_W^e$ and can have either sign for $Q_W^p$, with positive sign
being somewhat more likely than the negative sign. The potential magnitude of the
effects are considerably larger than those generated by SUSY loops. In
principle, then,
a comparison of $Q_W^e$ and $Q_W^p$ can potentially
establish whether or not R-parity is violated within a SUSY extension of
the SM. Having
an answer to this question would have consequences reaching beyond the realm
of accelerator physics.
For instance, if $P_R$ is violated in PVES, then lepton number is not conserved,
thereby implying that neutrinos
have Majorana masses and making neutrinoless
double beta decay possible (see {\it e.g.} Ref.~\cite{PDG00}). R-parity
violation also renders
the lightest supersymmetric particle unstable, thus eliminating SUSY dark
matter, which
has significant implications for cosmology \cite{kamion-cosm}.

Our discussion of these points is organized as follows.  After briefly
reviewing the
Minimal Supersymmetric Standard Model in
Section \ref {sec:theory}, we discuss the structure of the one-loop
radiative corrections to $Q_W^{e,p}$
in Section \ref{sec:general} and the tree-level $P_R$-violating
contributions in Section \ref{sec:rpv-contrib}.
The analysis of the prospective implications of the
parity-violating electron scattering measurements for supersymmetry is
presented in
Section \ref{sec:nc-analysis}. We conclude in Section \ref{sec:conclusion}.
Appendix \ref{app:counterterms} lists the counterterms for the effective
PVES Lagrangian of
Eq.~(\ref{eq:Leff}) necessary for renormalization of the one-loop radiative
corrections to the weak charges.
In Appendix \ref{sec:glue-dec}
we explicitly prove that gluino loops do not contribute to $Q_W^p$. In
Appendix \ref{app:expressions}
we give complete expressions for all process-dependent one-loop SUSY
corrections to the
$ep$ and $ee$ scattering; expressions for process-independent contributions
are given in the appendices of
Ref.~\cite{kur-rm-su-nutev}.

\section{MSSM Parameters}
\label{sec:theory}

The content of the MSSM has been described in detail
elsewhere\cite{haber-kane},
so we review only a
few features here. The particle spectrum consists of the SM particles and
the corresponding
superpartners: spin-0 sfermions ($\tilde f$, which include sneutrinos
$\tilde \nu$,
charged sleptons $\tilde l$, and
up- and down-type squarks $\tilde u$ and $\tilde d$), spin-$1/2$
gluinos ($\tilde g$), spin-$1/2$ mixtures of neutral Higgsinos ($\tilde
H_{1-2}^0$), the bino ($\tilde B$),
and the neutral wino ($\tilde W^3$), collectively called neutralinos
($\chi^0_{1-4}$), and spin-$1/2$ mixtures of charged Higgsinos
($\tilde H^\pm$) and
charged winos ($\tilde W^\pm$), collectively called charginos
($\chi^{\pm}_{1,2}$).
In addition, the
Higgs sector of
the MSSM contains two doublets (up- and down-types, which
give mass to the up- and down-type
fermions, respectively), whose
vacuum expectations $v_u$ and $v_d$ are parameterized in terms of
$v=\sqrt{v_u^2+v_d^2}$
and $\tan\beta=v_u/v_d$. Together with the SU(2)$_L$ and U(1)$_Y$ couplings
$g$ and $g'$
respectively, $v$ is determined from $\alpha$, $M_Z$, and $G_\mu$, the Fermi
constant extracted
from the muon lifetime, while $\tan\beta$ remains a free
parameter. The MSSM
also introduces a coupling between the two Higgs doublets characterized by the
dimensionful parameter $\mu$. The complete set Feynman rules for the MSSM,
which take
into account SUSY breaking and particle mixing, are given in Ref.
\cite{rosiek}.

Degeneracy between SM particles and their superpartners is lifted by the
SUSY-breaking
Lagrangian, which depends in general on 105 additional parameters. These
include the SUSY-breaking Higgs mass parameters; the
electroweak gaugino masses $M_{1,2}$; the gluino mass $M_{\tilde g}$; the left-
(right-)handed sfermion mass
parameters
$M_{\tilde f_L}^2$ ($M_{\tilde f_R}^2$); and left-right mixing terms
$M_{\tilde f_{LR}}^2$ which mix ${\tilde f}_L$ and
${\tilde f}_R$ into mass eigenstates ${\tilde f}_{1,2}$.
In our analysis, we take the sfermion mass matrices to be diagonal in
flavor space to avoid large flavor-changing
neutral currents. We also set all CP-violating phases to zero.
One expects the magnitude of the SUSY-breaking
parameters to lie
somewhere between the weak scale and $\sim 1$ TeV. Significantly larger
values can
reintroduce the hierarchy problem.

Theoretical models for SUSY-breaking mediation provide
relations among this large set of soft SUSY-breaking parameters, generally
resulting in only a few
independent parameters at the SUSY-breaking or GUT scale\cite{Kane}.
Evolution of the soft parameters down to the weak scale introduces
flavor- and species-dependence
into the superpartner spectrum due to the presence of Yukawa and gauge couplings in the renormalization
group (RG) equations. According to the model-independent  analysis of Ref.
\cite{kurylov02}, however,
generic features of this spectrum implied by typical SUSY-breaking models
and RG evolution may conflict
with the combined constraints of low-energy charged current data,
$M_W$, and the muon anomalous magnetic moment unless one allows for
non-conservation of $P_R$. In light of this situation, we adopt here a
similar model-independent approach and do
not impose any specific relations
among SUSY-breaking parameters. To our knowledge, no other
model-independent analysis of MSSM
corrections to PV observables has appeared in the literature,
nor have the complete
set of corrections to low-energy PV observables been computed previously
(see, {\it e.g.} Ref.~\cite{erler} for a study within minimal supergravity
and gauge mediated models of SUSY breaking).

\section{Radiative Corrections to $Q_W^f$}
\label{sec:general}

With higher-order corrections included, the weak charge of a fermion $f$
can be written as
\begin{equation}
\label{eq:C1f-radcorr}
Q_W^f = \rho_{PV}\left[2 I_3^f -4 Q_f\kappa_{PV}\sin^2\theta_W\right]+\lambda_f~,
\end{equation}
where $I_3^f$ and $Q_f$ are, respectively, the weak isospin and the
electric charge of the fermion $f$.
The quantities
$\rho_{PV}$ and $\kappa_{PV}$ are universal in that they do not depend on
the fermion $f$ under
consideration. The correction $\lambda_f$, on the other hand, does depend
on the
fermion species. At tree-level, one has $\rho_{PV}=1=\kappa_{PV}$ and
$\lambda_f=0$, while at one-loop order, these parameters are
\begin{eqnarray}
\label{eq:rho-kappa-lam}
\rho_{PV}&=&1+\delta \rho^{\rm SM} + \delta \rho^{\rm SUSY} ~,\nonumber \\
\kappa_{PV}&=&1+\delta \kappa^{\rm SM} + \delta \kappa^{\rm SUSY} ~,\nonumber \\
\lambda_f&=&\lambda_f^{\rm SM} + \lambda_f^{\rm SUSY}~,
\end{eqnarray}
where the SUSY contributions to $\rho_{PV}$, $\kappa_{PV}$, and $\lambda_f$
are denoted in the above equation by the corresponding superscript. In
general, the
corrections $\delta\rho$, $\delta\kappa$, {\em etc.} depend on
$q^2$, and
in particular, the $q^2$-dependence of $\kappa_{PV}$ defines the scale-dependence of
the weak mixing angle: $\sin^2\theta_W^{eff}(q^2) = \kappa_{PV}(q^2)
\sin^2\theta_W$, with $\sin^2\theta_W$ being evaluated at some
reference scale $q^2_0$ (usually $q^2_0=M_Z^2$).

\begin{figure}
\begin{center}
\resizebox{16. cm}{!}{\includegraphics*[0,430][640,730]{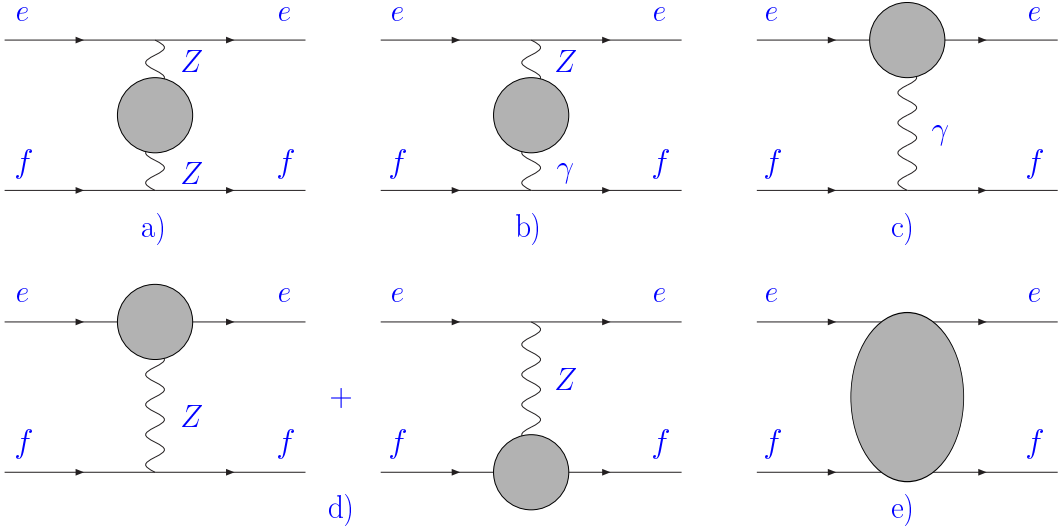}}
\caption{Types of radiative corrections to parity-violating electron
scattering:
(a) $Z$ boson
self-energy, (b) $Z-\gamma$ mixing, (c) electron anapole moment
contributions, (d) vertex corrections,
and (e) box graphs. External leg corrections are not explicitly shown.}
\label{fig:nc-general}
\end{center}
\end{figure}

The precise definitions of $\sin^2\theta_W$, $\kappa_{PV}(q^2)$, {\em etc.}
depend on one's
choice of renormalization scheme.
We evaluate the SUSY contributions using the modified dimensional reduction
renormalization scheme (${\overline{\rm DR}}$) \cite{siegel} and denote all
quantities
evaluated in this scheme by a hat. In ${\overline{\rm DR}}$, all
momenta are extended to $d=4-2\epsilon$ dimensions, while the Dirac algebra
remains four-dimensional
as required by SUSY invariance. The relevant classes of Feynman diagrams
are shown in Fig.~\ref{fig:nc-general}. Note that all gauge boson
self-energies contribute only to $\rho_{PV}$ and $\kappa_{PV}$
while all non-universal box diagrams as well as vertex and external leg
corrections are combined in $\lambda_f$. The counterterms for the effective
PVES
Lagrangian in Eq.~(\ref{eq:Leff}) in the ${\overline{\rm DR}}$ scheme are
given in Appendix \ref{app:counterterms}.

The $Z$ boson self-energy contribution (Fig.~\ref{fig:nc-general}a)
simply rescales the leading order amplitude. Its effect is naturally
combined with the counterterm $\delta {\hat G_\mu}$ from Eq.
(\ref{eq:Gmu-counter}) into $\rho_{PV}$:
\begin{equation}
\rho_{PV}=1+{{\delta {\hat G_\mu}} \over G_\mu}+{{\hat \Pi}_{ZZ}(0)\over M_Z^2}
=1-{{\hat \Pi}_{WW}(0)\over M_W^2}+{{\hat \Pi}_{ZZ}(0)\over M_Z^2}-{\hat
\delta}_{VB}^\mu~,
\end{equation}
where the $Z$ boson self-energy is evaluated at $q^2=0$. This is an
appropriate
approximation in this case because the momentum transfer in $ee$ and $ep$
scattering will be much smaller than the masses of the particles that
appear in the loop graphs. The error is of the order
$|q^2|/M_Z^2\sim 10^{-6}$, which is negligible. The quantity ${\hat
\delta}_{VB}^\mu$ denotes the
sum of electroweak vertex, external leg, and box graph corrections to the muon
decay amplitude, which
must be subtracted when the neutral current (NC) amplitudes are normalized to $G_\mu$.

The graphs Fig.~\ref{fig:nc-general}b,c contribute to $\kappa_{PV}$
in Eq.~(\ref{eq:C1f-radcorr}). The expression is:
\begin{equation}
\label{eq:kappa}
\kappa_{PV}=1+{{\hat c}\over {\hat s}}
{{\hat \Pi}_{Z\gamma}(q^2)\over q^2}+4{\hat c}^2 F_A^e(q^2)+{\delta{\hat
s}^2\over {\hat s}^2}~,
\end{equation}
where $F_A^e(q^2)$ is the parity-violating electron-photon
form factor, which -- at $q^2=0$ -- is known as the anapole moment of the electron (see
Eq.~(\ref{eq:nc-anapole}) below)
\footnote{
Note that our definition of $\kappa_{PV}$, Eq.~(\ref{eq:kappa}), includes the
anapole form-factor of the electron $F_A^e(q^2)$, which may be absent in
other definitions appearing in literature (see {\it e.g.} Ref.~\cite{Erl-MJRM-Kur02}).
}.
It should be noted that in the MSSM one has
${\hat \Pi}_{Z\gamma}^{\rm SUSY}(q^2)\sim q^2$, so there is no singularity
at $q^2=0$ in the above equation \cite{pierce}. The SM contribution contains a
singularity that is canceled by a corresponding singularity in the anapole
moment contribution
(Fig.~\ref{fig:nc-general}c). Since in the following we consider only the
new physics contributions,
this issue is irrelevant (a complete treatment of the SM
contributions is given in Ref.~\cite{sirlin-nc,Mar96,Erl-MJRM-Kur02}). The
shift $\delta{\hat s}^2$ in ${\hat s^2}=1-{\hat c^2}\equiv\sin^2{\hat\theta}_W(M_Z^2)$
arises from its definition in terms of $\alpha$, $G_\mu$, and $M_Z$:
\begin{eqnarray}
\label{eq:s2-dr-bar}
{\hat s}^2{\hat c}^2&=&{\pi\alpha\over{{\sqrt 2}G_\mu M_Z^2}
(1-\Delta {\hat r})} ~, \nonumber \\
\Delta {\hat r}&=&{{\hat\Pi}^\prime}_{\gamma\gamma}(0)+2{{\hat s}\over
{\hat c}}
{{\hat \Pi}_{Z\gamma}(0)\over M_Z^2}-{{\hat \Pi}_{ZZ}(M_Z^2)\over M_Z^2}
+{{\hat \Pi}_{WW}(0)\over M_W^2}+{\hat\delta}_{VB}^\mu
~,\end{eqnarray}
where ${{\hat\Pi}^\prime}_{\gamma\gamma}(q^2)\equiv{\hat
\Pi}_{\gamma\gamma}/q^2$.
Writing ${\Delta \hat r}={\Delta \hat r}^{\rm SM}+{\Delta \hat r}^{\rm SUSY}$
one has
\begin{equation}
\label{eq:s2-sm-expr}
{\delta{\hat s}^2_{\rm SUSY}\over {\hat s}^2}={{\hat c}^2\over{{\hat
c}^2-{\hat s}^2}} \Delta{\hat r}^{\rm SUSY}
~.\end{equation}
In computing the SUSY corrections to the weak charges one must decide which
value for
${\hat s}^2$ to use. Since $\delta{\hat s}^2_{\rm SUSY}$ has already been
absorbed into $\kappa_{PV}$
one must determine ${\hat s}^2$ from Eq.~(\ref{eq:s2-dr-bar}) using the SM
radiative corrections only.
The corresponding value extracted using only $\alpha$, $G_\mu$, and $M_Z$
is \cite{PDG00}:
\begin{equation}
{\hat s}^2=0.23120\pm0.00018~.
\end{equation}

In order to incorporate constraints from existing precision data (see
Section \ref{sec:nc-analysis}), it is
useful to introduce the oblique parameters $S$, $T$, and $U$ \cite{stu-degrassi}:
\begin{eqnarray}
\label{eq:stu-sirlin}
S&=&{4{\hat s}^2{\hat c}^2\over {{\hat \alpha}M_Z^2}}{\rm Re}\Biggl\{
{\hat \Pi}_{ZZ}(0)-{\hat \Pi}_{ZZ}(M_Z^2)+{{\hat c}^2-{\hat s}^2\over{\hat
c}{\hat s}}
\left[{\hat \Pi}_{Z\gamma}(M_Z^2)-{\hat \Pi}_{Z\gamma}(0)\right]
+{\hat \Pi}_{\gamma\gamma}(M_Z^2)
\Biggr\}^{\rm New} ~,\nonumber \\
T&=&{1\over {{\hat \alpha}M_W^2}}
\Biggl\{
{\hat c}^2\left( {\hat \Pi}_{ZZ}(0)+{2{\hat s}\over
{\hat c}}{\hat \Pi}_{Z\gamma}(0) \right)
-{\hat \Pi}_{WW}(0)
\Biggr\}^{\rm New} ~,\nonumber \\
U&=&{4{\hat s}^2\over {{\hat \alpha}}}
\Biggl\{
{{\hat \Pi}_{WW}(0)-{\hat \Pi}_{WW}(M_W^2)\over M_W^2}
+{\hat c}^2{{\hat \Pi}_{ZZ}(M_Z^2)-{\hat \Pi}_{ZZ}(0)\over M_Z^2} \nonumber \\
&+&2{\hat c}{\hat s}{{\hat \Pi}_{Z\gamma}(M_Z^2)-{\hat \Pi}_{Z\gamma}(0)\over M_Z^2}
+{\hat s}^2 {{\hat \Pi}_{\gamma\gamma}(M_Z^2)\over M_Z^2}
\Biggr\}^{\rm New}
~,\end{eqnarray}
where the superscript \lq\lq New" indicates that only the new physics
contributions
to the self-energies are included.
Contributions to gauge-boson self energies  can be expressed entirely in
terms of the oblique parameters $S$, $T$, and $U$ in the limit that
$M_{\rm NEW}\gg M_Z$. However, since present collider limits allow for
fairly light superpartners,
we do not work in this limit. Consequently, the corrections arising from
the photon
self-energy ($\Pi_{\gamma\gamma}$) and $\gamma$-$Z$ mixing tensor
($\Pi_{Z\gamma}$)
contain a residual $q^2$-dependence not embodied by the oblique parameters.
Expressing
$\rho_{PV}$ and $\kappa_{PV}$ in terms of $S$,$T$, and $U$ we obtain:
\begin{eqnarray}
\label{eq:rho-kappa-stu}
\delta\rho^{\rm SUSY} & = & {\hat\alpha} T-{\hat\delta}_{VB}^\mu ~,\nonumber \\
\delta\kappa^{\rm SUSY} & = &
\left({{\hat c}^2\over {\hat c}^2-{\hat s}^2} \right)
\left({{\hat\alpha}\over 4{\hat s}^2 {\hat c}^2} S-{\hat \alpha} T
+{\hat\delta}_{VB}^\mu \right)
+{{\hat c}\over {\hat s}}\Bigl[  {{\hat\Pi}_{Z\gamma}(q^2)\over q^2}-
{{\hat\Pi}_{Z\gamma}(M_Z^2)\over M_Z^2}\Bigr]^{\rm SUSY} \nonumber \\
&&+\Bigl({{\hat c}^2\over {\hat c}^2-{\hat s}^2}
\Bigr)\Bigl[-{{\hat\Pi}_{\gamma\gamma}(M_Z^2)\over M_Z^2}
+{\Delta{\hat\alpha}\over \alpha}
\Bigr]^{\rm SUSY}+4{\hat c}^2 F_A^{e}(q^2)^{\rm SUSY}
~,\end{eqnarray}
where $\Delta{\hat\alpha}$ is the SUSY contribution to the difference
between the
fine structure constant and the electromagnetic coupling renormalized
at $\mu=M_Z$:
$\Delta {\hat \alpha}=\left[{\hat \alpha}(M_Z)-\alpha\right]^{\rm SUSY}$.
As noted above, we take $q^2\to 0$ in our analysis.

The non-universal contribution to the weak charge is determined by the sum
of the renormalized vertex
corrections ${\hat V}_{V,A}^f$ in Fig.~\ref{fig:nc-general}d [see Eq.~(\ref{eq:vert-renorm})]
and the box graphs ${\hat \delta}_{\rm Box}^{ef}$
in Fig.~\ref{fig:nc-general}e [see Eq.~(\ref{eq:del-box})] :
\begin{equation}
\label{eq:nc-lambda}
{\hat\lambda}_f=g_V^f {\hat V}_A^e+g_A^e {\hat V}_V^f+{\hat \delta}_{\rm Box}^{ef}~,
\end{equation}
where $g_{V,A}^f$ are given in Eq.~(\ref{eq:gs}).

Finally, we note that by vector current conservation, $\delta Q_W^p$ can be
computed directly from the shifts in
the up- and down-quark weak charges: $\delta Q_W^p=2 \delta Q_W^u+\delta
Q_W^d$. The analogous relation
in the SM is modified by non-perturbative strong interactions in the
$Z\gamma$ box graph\cite{Erl-MJRM-Kur02}.
The latter arise because the loop contains a massless particle, rendering
the corresponding loop integral
sensitive to both low and high momentum scales.
In contrast, the SUSY radiative corrections are dominated by large loop
momenta, and non-perturbative QCD
corrections are suppressed by  $(\Lambda_{\rm QCD}/M_{\rm SUSY})^2\ll 1$.

\subsection*{One-Loop SUSY Feynman Diagrams}
\label{sec:susy-loop}

Here, we present the SUSY one-loop diagrams that are particular to
PVES. Such diagrams
correspond to the generic corrections shown in Fig.
\ref{fig:nc-general}c,d, and e. Contributions corresponding to
Fig.~\ref{fig:nc-general}a and b are universal, and the relevant diagrams
-- together with the external leg
corrections for all fermions -- are given in Ref.~\cite{kur-rm-su-nutev}.
In addition, some simplifications occur in the analysis
for PVES that do not arise
in general. In the case of charged current observables, for example, gluino
loops can generate
substantial corrections \cite{kurylov02,kur-rm-su-nutev}. In contrast,  gluinos decouple entirely
from the
one-loop MSSM corrections to semi-leptonic neutral current PV observables.
The proof of this statement
is given in Appendix \ref{sec:glue-dec}.
In addition, the MSSM Higgs contributions to
vertex, external leg, and box graph corrections are
negligible due to the small, first- and second-generation Yukawa couplings.
The light Higgs
contribution to gauge boson propagators has already been included via the
oblique parameters,
while the effects of other MSSM Higgs bosons are sufficiently small to be
neglected\cite{haber93}. Therefore, we do not discuss gluino and Higgs
contributions in the following.

{\bf Anapole moment corrections, corresponding to Fig.
\ref{fig:nc-general}(c).}
In the presence of
parity-violating interactions, higher-order contributions can generate the
photon-fermion
coupling of the form (see {\it e.g.} Ref.~\cite{rm-anapole}):
\begin{equation}
\label{eq:nc-anapole}
i{\cal M}_{\gamma-f}^{PV}=-ie {F_A^f(q^2)\over M_Z^2}{\bar f}(q^2
\gamma^\mu-{\slash q}q^\mu)
\gamma_5 f \varepsilon_\mu
~,\end{equation}
where $f$ is a fermion spinor, $\varepsilon_\mu$ is the photon polarization, and
$F_A^f(q^2)$ is the anapole moment form-factor.

The quantity $F_A^f(q^2)$ is, in general, gauge-dependent. This dependence
cancels after the anapole moment contribution
is combined with other one-loop corrections to the given scattering process
\cite{rm-anapole}.
The Feynman diagrams that contribute to $F_A^e(q^2)^{\rm SUSY}$ are shown in
Fig \ref{fig:nc-anapole}, and the
analytical expressions are presented in Appendix \ref{app:expressions}.
\begin{figure}[ht]
\hspace{0.00in}
\begin{center}
\resizebox{10.
cm}{!}{\includegraphics*[130,600][470,770]{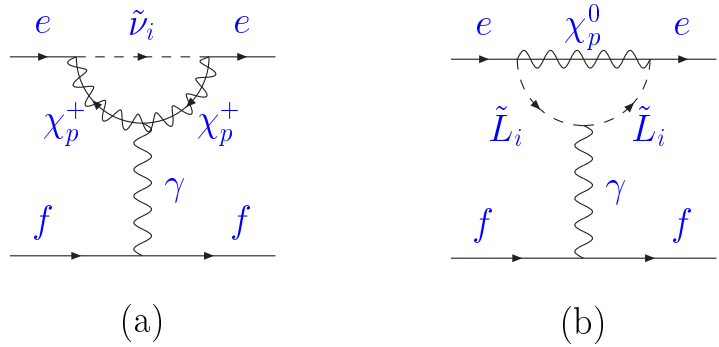}}
\caption{Electron anapole moment contributions to the parity-violating
electron-fermion scattering amplitude.}
\label{fig:nc-anapole}
\end{center}
\end{figure}

\begin{figure}
\hspace{0.00in}
\begin{center}
\resizebox{16.
cm}{!}{\includegraphics*[40,610][590,770]{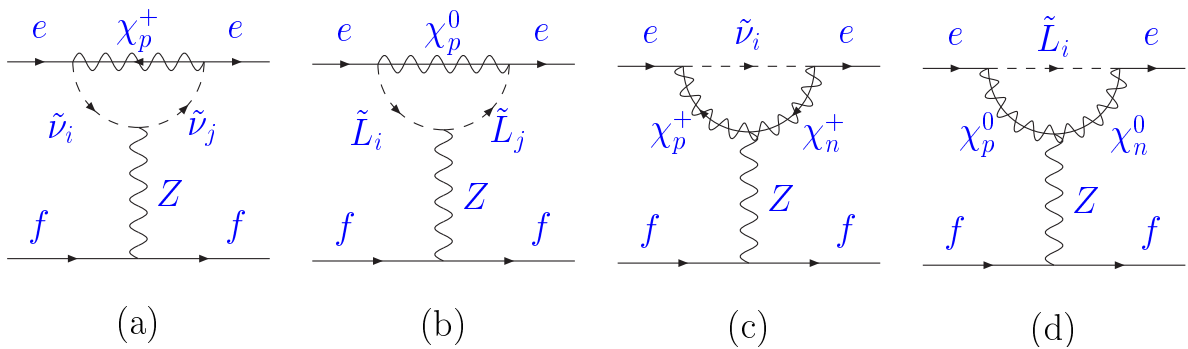}}
\caption{MSSM radiative corrections to the electron neutral current vertex.
Radiative corrections to
the down quark vertex are obtained by replacing charged leptons (sleptons)
with down type quarks (squarks) and sneutrinos with up type squarks.}
\label{fig:nc-electron-vertex}
\end{center}
\end{figure}

\begin{figure}
\hspace{0.00in}
\begin{center}
\resizebox{16.
cm}{!}{\includegraphics*[40,610][590,770]{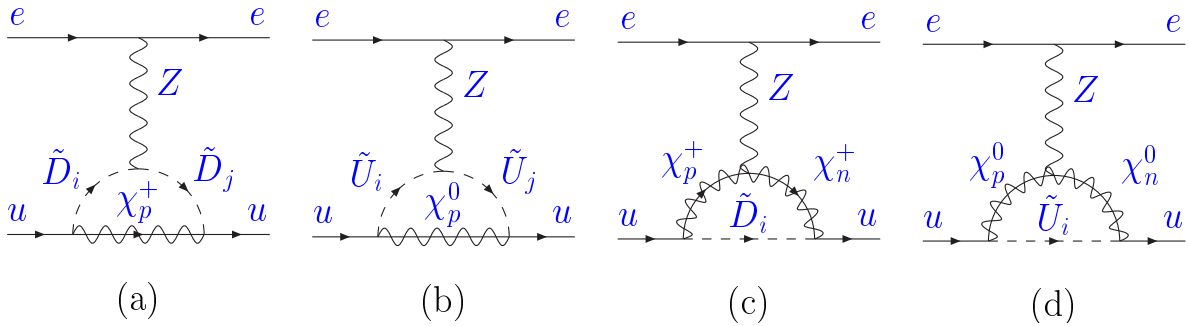}}
\caption{MSSM radiative corrections to the up quark neutral current vertex.}
\label{fig:nc-uquark-vertex}
\end{center}
\end{figure}

{\bf Vertex corrections, corresponding to Fig.~\ref{fig:nc-general}(d).}
The relevant diagrams are shown in Figs. \ref{fig:nc-electron-vertex}
and \ref{fig:nc-uquark-vertex}.
The diagrams in Fig.~\ref{fig:nc-electron-vertex} cover $ee$, $ed$, and $eu$
scattering when the radiative correction is for the projectile side.
When the radiative correction is to be applied to the target side,
the diagrams in Fig.~\ref{fig:nc-electron-vertex} can also be used
for $ee$ and $ed$ scattering. In this case $f=e$ is the projectile.
To obtain the corrections to the down quark vertex, the electron can simply be
replaced with the down quark. The diagrams in Fig.~\ref{fig:nc-uquark-vertex}
show the radiative corrections to the target side when the incoming
electron interacts with the up quark inside the proton.
The explicit expressions for the vertex corrections can be found in
Appendix \ref{app:expressions}.

{\bf Box corrections, corresponding to Fig.~\ref{fig:nc-general}(e).}
These graphs generate ${\hat\delta}_{\rm Box}^{ef}$ in Eq.~(\ref{eq:nc-lambda}).
The relevant diagrams are shown in Fig.
\ref{fig:nc-electron-boxes} and Fig.~\ref{fig:nc-uquark-boxes}.
The explicit expressions are
given in Appendix \ref{app:expressions}.
\begin{figure}[ht]
\hspace{0.00in}
\begin{center}
\resizebox{16.
cm}{!}{\includegraphics*[0,580][640,730]{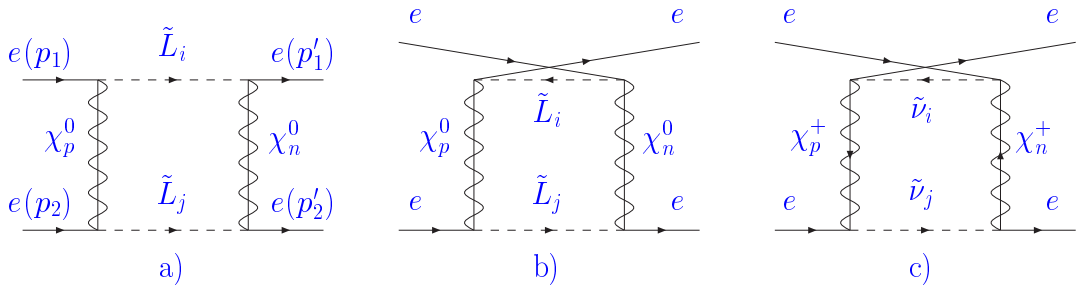}}
\caption{MSSM box graphs that contribute to the electron-electron
scattering amplitude.
Here, $p_i$($p_i^\prime$), $i=1,2$, is the momentum of the initial (final)
state fermion.
Radiative corrections to the electron-down quark scattering are trivially
obtained by
replacing the target with the down quark.}
\label{fig:nc-electron-boxes}
\end{center}
\end{figure}
\begin{figure}[ht]
\hspace{0.00in}
\begin{center}
\resizebox{16.
cm}{!}{\includegraphics*[0,580][640,730]{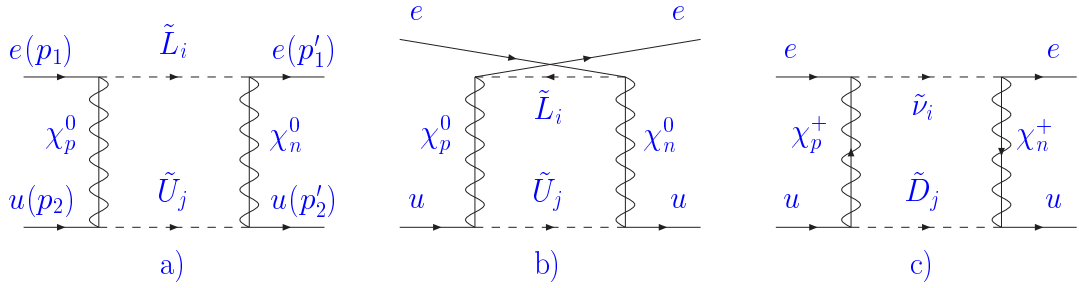}}
\caption{MSSM box graphs that contribute to the electron-up quark scattering.
The meaning of momentum labels is the same as in Fig.
\ref{fig:nc-electron-boxes}.}
\label{fig:nc-uquark-boxes}
\end{center}
\end{figure}

\section{R-parity violating contributions to $Q_W^{\MakeLowercase f}$}
\label{sec:rpv-contrib}

\begin{figure}[ht]
\begin{center}
\resizebox{16. cm}{!}{\includegraphics*[20,580][600,720]{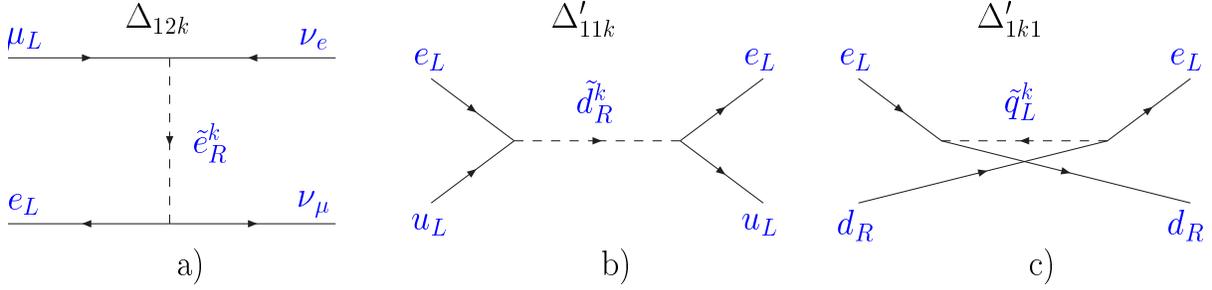}}
\caption{
Tree-level $P_R$-violating contributions to the muon decay [plot (a)], the
$eu$ scattering
amplitude [plot (b)], and the $ed$ scattering amplitude [plot
(c)]. The quantities $\Delta_{ijk}$, {\it etc.}, 
are defined in Eq.~(\ref{eq:deltas}).
}
\label{fig:rpv-graphs}
\end{center}
\end{figure}

When R-parity is not conserved, new tree-level contributions to $Q_W^{e,p}$
appear. The latter are generated by the
$(B-L)$-violating superpotential:

\begin{equation}
\label{eq:RPV-super}
W_{\sst{RPV}} = \frac{1}{2}\lambda_{ijk}L_i L_j {\bar E}_k
+\lambda^{\prime}_{ijk}L_i Q_j{\bar D}_k +
\frac{1}{2}\lambda^{\prime\prime}_{ijk}{\bar U}_i{\bar D}_j{\bar
D}_k+\mu^{\prime}_{i}L_iH_u\ \ \ ,
\end{equation}
where $L_i$ and $Q_i$ denote lepton and quark SU(2)$_L$ doublet
superfields, $E_i$, $U_i$, and $D_i$ are
singlet superfields and the $\lambda_{ijk}$ {\em etc.} are {\em a priori}
unknown couplings. In order to
avoid unacceptably large contributions to the proton decay rate, we set the
$\Delta B\not= 0$ couplings
$\lambda^{\prime\prime}_{ijk}$ to zero.
For simplicity, we also neglect the last term in Eq.~(\ref{eq:RPV-super}).
The purely leptonic terms
($\lambda_{12k}$) contribute to the electron scattering
amplitudes via the normalization of NC amplitudes to $G_\mu$ and
through the definition of
$\hat s^2$\cite{MRM00}.  The remaining semileptonic, $\Delta L=\pm 1$
interactions ($\lambda^{\prime}_{ijk}$) give
direct contributions to the $eq$ scattering amplitudes. The latter may be
obtained computing the Feynman amplitudes in Fig.~\ref{fig:rpv-graphs}(b,c)
and preforming a Fierz reordering. In this manner one obtains the following
effective four-fermion Lagrangian:
\begin{eqnarray}
\label{eq:rpveffective}
{\cal L}_{\sst{RPV}}^{\sst{EFF}} & = & -{|\lambda^{\prime}_{1k1}|^2\over 2
M^2_{\tilde q^k_L}}{\bar d}_R\gamma^\mu d_R
{\bar e}_{L}\gamma_\mu e_{L} +
{|\lambda^{\prime}_{11k}|^2\over 2 M^2_{\tilde d^k_R}}{\bar u}_L\gamma^\mu
u_L {\bar e}_{L}\gamma_\mu e_{L} \nonumber \\
&&  - {|\lambda_{12k}|^2\over 2 M^2_{\tilde e^k_R}}\Biggl[{\bar
\nu}_{\mu L}\gamma^\mu \mu_L
{\bar e}_{ L}\gamma_\mu \nu_{e L}+{\rm h.c.}\Biggr]\ \ \ ,
\end{eqnarray}
where we have taken $|q^2|\ll M_{\tilde f}^2$ and have retained only the
terms relevant for the PVES scattering. Note the absence from Eq.
(\ref{eq:rpveffective}) of the parity-violating
contact four-electron interaction. It is straightforward to show that the
superpotential in
Eq.~(\ref{eq:RPV-super}) can only produce parity-conserving contact
interactions between identical
leptons.

Contributions from $P_R$-violating
interactions to low energy observables can be parametrized in terms of the
following quantities:
\begin{equation}
\label{eq:deltas}
\Delta_{ijk}(\tilde f)={|\lambda_{ijk}|^2\over 4\sqrt{2}G_\mu M_{\tilde
f}^2}\ge 0
~,\end{equation}
with a similar definition for the primed quantities.
In terms of $\Delta_{ijk}$ {\em etc.} one obtains for the relative shifts
in the weak charges \cite{MRM00}:
\begin{eqnarray}
\label{eq:rpv-weak}
{\delta Q_W^{e}\over Q_W^{e}}&\approx&-\left[1+
\left({4\over 1-4\sin^2\theta_W}\right)\lambda_x
\right]\Delta_{12k}({\tilde e}_R^k) ~,\nonumber \\
{\delta Q_W^{p}\over Q_W^{p}}&\approx &\left({2\over 1-4\sin^2\theta_W}\right)
\left[
-2\lambda_x \Delta_{12k}({\tilde e}_R^k) +2\Delta_{11k}^\prime({\tilde
d}_R^k)-\Delta_{1k1}^\prime({\tilde q}_L^k)\right]-\Delta_{12k}({\tilde e}_R^k)~,\nonumber \\
\lambda_x&=&{{\hat s}^2(1-{\hat s}^2)\over 1-2{\hat s}^2}
{1\over 1-\Delta {\hat r^{\rm SM}}} \approx 0.35
~.\end{eqnarray}
As discussed in Section \ref{sec:nc-analysis} the quantities
$\Delta_{ijk}$ {\em etc.} are constrained from
other precision data. Since they are non-negative, Eq.~(\ref{eq:rpv-weak})
indicates that the relative
shift in $Q_W^e$ is negative semidefinite. On the other hand, the relative shift
in $Q_W^p$ can have either sign depending on the relative magnitudes
of $\Delta_{12k}$, $\Delta_{11k}^\prime$, and $\Delta_{1k1}^\prime$.

\section{Analysis of the SUSY contributions to the weak charges}
\label{sec:nc-analysis}

In order to evaluate the potential size of SUSY loop corrections,
a set of about 3000 different combinations of SUSY-breaking
parameters was generated, chosen randomly from a flat distribution in
the soft SUSY mass parameters (independent for each generation) and in
$\ln\tan\beta$. The former were
bounded below by present collider limits
and bounded above by 1000 GeV, corresponding to the ${\cal O}$(TeV)
naturalness limit. Also, $\tan\beta$ was restricted to lie in the
range $1.4 < \tan\beta < 60$. These limits follow from the requirement
that the third generation quark Yukawa couplings remain
perturbative (small)  up to the GUT scale.
Left-right mixing among sfermions was allowed. In order to avoid unacceptably
large flavor-changing neutral currents, no
intergenerational sfermion mixing was permitted. The ranges over which
the soft SUSY breaking parameters and $\tan\beta$ were scanned are shown in Table~\ref{table:scanparams}.
\begin{table}[th]
\begin{tabular}{|c|cc|}
\hline
Parameter & Min & Max \\
\hline
\hline
$\tan\beta$ & 1.4 & 60 \\
${\tilde M}$ & 50 GeV & 1000 GeV \\
{\rule[-10pt]{0pt}{15pt}$\left(M^2_{\tilde f}\right)^i_{LR}$}
& -$10^6~{\rm GeV}^2$ & $10^6~{\rm GeV}^2$ \\
\hline
\end{tabular}
\caption{Ranges of SUSY parameters scanned.  Here, ${\tilde M}$ denotes
any of $|\mu|$, $M_{1,2}$, or
the diagonal sfermion mass parameters
$M^i_{{\tilde f}_{L,R}}$. The $\mu$ parameter and $M_{1,2}$ can
take either sign.  The generation index $i$ runs from 1 to 3.}
\label{table:scanparams}
\end{table}

For each combination of parameters, we computed superpartner masses and
mixing angles, which we
then used as inputs for computing the radiative corrections. Only the
parameters generating
SUSY contributions to the muon anomalous magnetic moment consistent with
the latest results
\cite{bnl-muon} were considered. We also separately evaluated the
corresponding contributions to the oblique parameters. The latter are
tightly constrained from
precision electroweak data.
We rule out any parameter combination leading
to values of $S$ and
$T$ lying outside the present 95\% confidence limit contour for these
quantities. We note that
this procedure is not entirely self-consistent, since we have not evaluated
non-universal MSSM
corrections to other precision electroweak observables before extracting
oblique parameter
constraints. As noted in Ref.
\cite{erler}, where MSSM corrections to $Z$-pole observables were evaluated
using
different models for SUSY-breaking mediation, non-universal effects can be
as large as
oblique corrections. Nevertheless, we expect our procedure to yield a
reasonable estimate
of the oblique parameter constraints.  Since  $S$ and $T$ do not dominate
the low-energy
SUSY corrections (see below), our results depend only gently on the precise
allowed ranges for
these parameters.

\begin{figure}[ht]
\begin{center}
\includegraphics[width=4in]{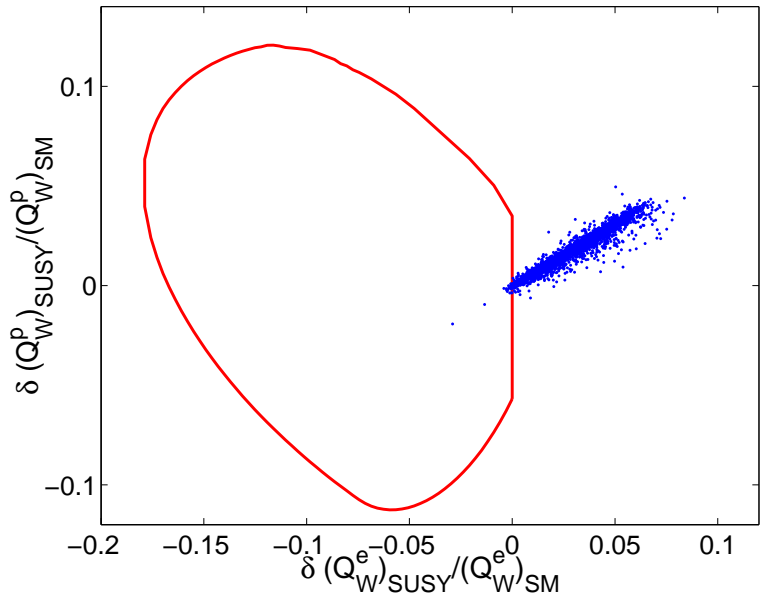}
\caption{Relative shifts in electron and proton weak charges
due to SUSY effects. Dots indicate MSSM loop corrections for $\sim 3000$
randomly-generated SUSY-breaking parameters. Interior of truncated
elliptical region gives possible shifts
due to $P_R$ nonconserving SUSY interactions (95\% confidence).}
\label{fig:mssm-vs-parity}
\end{center}
\end{figure}

In Fig.~\ref{fig:mssm-vs-parity}, we plot the shift in the weak charge
of the proton, $\delta Q_W^p = 2\delta Q_W^u+ \delta Q_W^d$, versus
the corresponding shift in the electron's weak
charge, $\delta Q_W^e$, normalized to the respective SM values.
The corrections in the
MSSM (with
$P_R$ conserved) can be as large as
$\sim 4\%$ ($Q_W^p$) and
$\sim 8\%$ ($Q_W^e$) -- roughly the size of the proposed experimental
errors for the two PVES measurements.
Generally speaking, the magnitudes of $\delta Q_W^{e,p}$ slowly increase
with $\tan\beta$ and decrease as SUSY
mass parameters are
increased. The largest effects occur when at least one superpartner is
relatively light.
An exception occurs in the presence of significant mass splitting between
sfermions, which may lead to sizable contributions. However, such weak
isospin-breaking
effects also increase the magnitude of $T$, so their impact is bounded by
oblique
parameter constraints. This consideration has been implemented in
arriving at Fig.~\ref{fig:mssm-vs-parity}.

The effects of sfermion left-right mixing were studied separately.
We observe that the presence
or the absence of the mixing
affects the distribution of points, but does not significantly change the
range of
possible corrections. For the situation of no left-right mixing, the points
are more
strongly clustered near the origin. Thus, while corrections of the order of
several percent
are possible in either case, large effects are more likely in the presence of
left-right mixing.
\begin{figure}[ht]
\begin{center}
\includegraphics[width=4in]{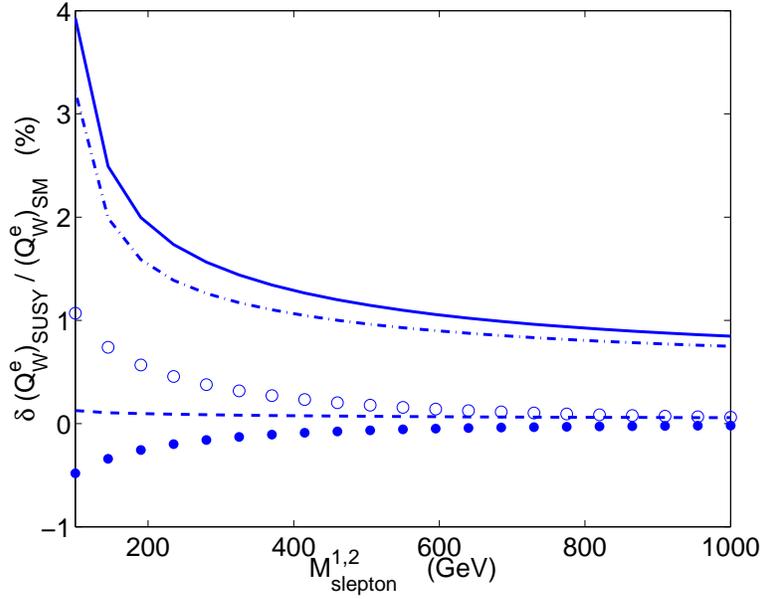}
\caption{Various contributions to $\delta Q_W^e/Q_W^e$: total from SUSY
loops (solid line),
from $\delta \rho^{\rm SUSY}$ (dashed line), from $\delta \kappa^{\rm
SUSY}$ (dash-dotted line), from the
vertex corrections (dotted line), and from the box graphs (open circles).
The $x$ axis gives the lepton superpartner mass, chosen to be the same for
both left- and right-handed
first and second generation sleptons.
For this graph, $\tan\beta=10$,
the gaugino soft mass parameters are $2M_1=M_2=\mu$ = 200 GeV, and masses of
the third
generation sleptons and of all squarks are 1000 GeV.
For this case, $\delta Q_W^e/Q_W^e$ is about half of its maximum possible value.
The total relative correction is clearly dominated by $\delta\kappa^{\rm SUSY}$.}
\label{fig:qwe-totals}
\end{center}
\end{figure}

The shifts $\delta Q_W^{e,p}$ are dominated by $\delta\kappa^{\rm
SUSY}$. This feature is illustrated for  $Q_W^e$ in Fig.~\ref{fig:qwe-totals}
where the soft SUSY breaking
parameters are chosen such that the total SUSY correction to $Q_W^e$ is
about 4\%, about a
half of its maximum value. For $Q_W^p$
the situation is similar.
We observe that non-universal
corrections involving vertex corrections and wavefunction renormalization
experience significant
cancellations.
In addition, corrections to $Q_W^{e,p}$ due to shifts in the $\rho_{PV}$
parameter are
suppressed by $1-4{\hat s}^2$.

We find that
$\delta\kappa^{\rm SUSY}$ is nearly always negative, corresponding to
a reduction in the  value of
$\sin^2{\theta}_W^{eff}(q^2)=\kappa_{PV}(q^2)\sin^2{\theta}_W$
for the parity-violating electron scattering experiments
[see Eq.~(\ref{eq:C1f-radcorr})].
In this case, the degree of cancellation between $2I^3_f$ and $Q_f$ terms
in Eq.~(\ref{eq:C1f-radcorr}) is reduced, yielding
an increased magnitude of $Q_W^f$. Since this effect is identical for both
$Q_W^e$ and $Q_W^p$, the dominant effect of
$\delta \kappa$ produces a linear correlation between the two weak charges.
Some scatter around this line arises
from non-universal effects in ${\hat\lambda}_f$ (see Fig.~\ref{fig:mssm-vs-parity}).

\begin{figure}[ht]
\begin{center}
\includegraphics[width=4in]{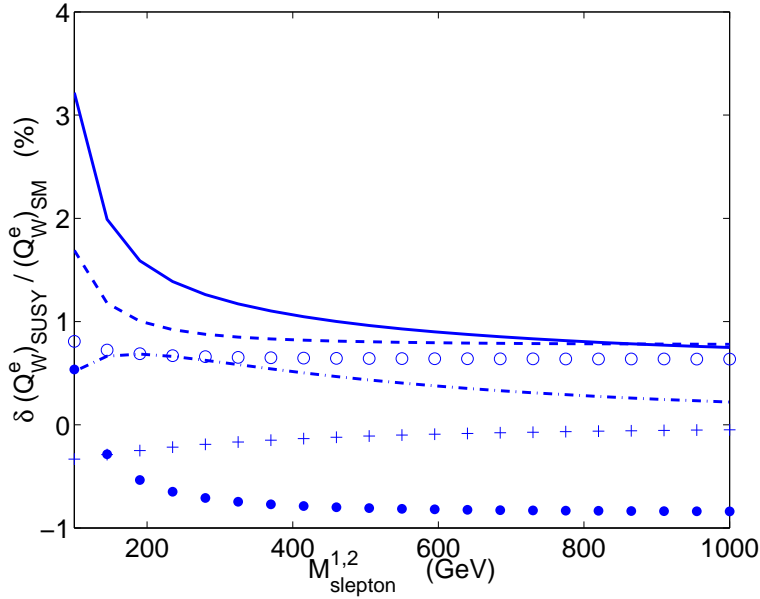}
\caption{Contributions to $\delta Q_W^{e}/Q_W^{e}$ from various corrections
to $\delta\kappa^{\rm SUSY}$ [see
Eq.~(\ref{eq:rho-kappa-stu})]: total from $\delta\kappa^{\rm SUSY}$ (solid
line), $S$ parameter (dotted line),
$T$ parameter (dashed line), ${\hat \delta}^\mu_{VB}$ (dash-dotted line), $Z-\gamma$
mixing and the photon self-energy (open circles), and the electron anapole moment (crosses). The soft
SUSY parameters are the same as in Fig.~\ref{fig:qwe-totals}.
}
\label{fig:qwe-kappa}
\end{center}
\end{figure}

As illustrated in Fig.~\ref{fig:qwe-kappa}, within $\delta\kappa^{\rm
SUSY}$ itself, contributions from the
various terms in Eq.~(\ref{eq:rho-kappa-stu}) have
comparable importance, with some degree of cancellation occurring between
the effects
of $S$ and $T$.
Thus, the oblique parameter approximation gives a rather poor description
of the
MSSM effects on the
weak charges. In particular the quantity ${\hat \delta}_{VB}^\mu$
in Eq.~(\ref{eq:rho-kappa-stu}) makes a significant contribution to
$\delta\kappa^{\rm SUSY}$.

As evident from Fig.~\ref{fig:mssm-vs-parity}, the relative sign of
the corrections to both $Q_W^p$ and
$Q_W^e$ -- normalized to the corresponding SM values -- is nearly
always the same and nearly always positive. Since $Q_W^p>0$
($ Q_W^e<0$) in the SM, SUSY loop corrections give $\delta Q_W^p>0$ ($\delta
Q_W^e<0$).
This correlation is significant, since the effects of other new
physics scenarios can display different signatures. For example, for the
general class of of theories based on $E_6$ gauge group, with neutral gauge
bosons having mass $\lsim$ 1000 GeV, the effects on
$Q_W^p$ and $Q_W^e$ also correlate, but $\delta Q_W^{e,p}/Q_W^{e,p}$ can
have either sign in this
case\cite{MRM99,Erl-MJRM-Kur02}. In contrast, leptoquark interactions would not lead
to discernible
effects in $Q_W^e$ but could induce sizable shifts in $Q_W^p$ \cite{Erl-MJRM-Kur02,MRM99}.

As a corollary, we also find that the relative importance of SUSY loop
corrections to
the weak charge of heavy nuclei probed with APV is suppressed. The shift in the
nuclear weak charge is given by $\delta Q_W(Z,N) = (2Z+N)\delta Q_W^u +
(2N+Z)\delta
Q_W^d$. Since the  sign of $\delta Q_W^f/Q_W^f$ due to superpartner loops
is nearly
always the same, and since $Q_W^u>0$ and $Q_W^d<0$ in the SM, a strong
cancellation between $\delta Q_W^u$ and $\delta Q_W^d$ occurs in heavy
nuclei. This
cancellation implies that the magnitude of $\delta Q_W(Z,N)/Q_W(Z,N)$ is
generally less than
about 0.2\% for cesium and is equally likely to have either sign. Since the
presently quoted
uncertainty for the cesium nuclear weak charge is
about 0.6\% \cite{sushov02}, cesium APV does not substantially constrain
the SUSY parameter space.
Equally as important, the present agreement
of $Q_W^{\rm Cs}$ with the SM prediction does not preclude significant shifts
in $Q_W^{e,p}$ arising from SUSY.
The situation is rather different, for example, in the $E_6~Z^\prime$
scenario, where sizable shifts in $Q_W^{e,p}$ would also imply observable
deviations of $Q_W^{\rm Cs}$ from the SM prediction.

The prospective \lq\lq diagnostic power" of the two PVES measurements is further
increased when one relaxes
the assumption of $P_R$ conservation. Doing
so leads to the tree-level corrections to the weak charges shown in Eq.~(\ref{eq:rpv-weak}).
The quantities $\Delta_{ijk}$, {\em etc.} in Eqs.~(\ref{eq:deltas}) and (\ref{eq:rpv-weak}) are
constrained from the existing precision
data
\cite{MRM00}. A summary of the existing constraints -- including the
latest theoretical inputs
into the extraction of $Q_W^{\rm Cs}$ from experiment \cite{sushov02} --  is
given in Table II of Ref.~\cite{kur-rm-su-nutev}, which we partially reproduce here in Table
\ref{tab:rpv-constr}. We list the $P_R$-violating
contribution to four relevant precision observables: superallowed  nuclear
$\beta$-decay that
constrains $|V_{ud}|$ \cite{towner-super}, atomic PV measurements of
the cesium weak charge $Q_W^{\rm Cs}$ \cite{Ben99}, the $e/\mu$ ratio
$R_{e/\mu}$ in $\pi_{l2}$ decays \cite{pil2}, and a comparison of the
Fermi constant $G_\mu$ with the appropriate combination of
$\alpha$, $M_Z$, and $\sstw$ \cite{marciano99}. The values of the experimental
constraints on those
quantities are given  in the last column. We also list the the $P_R$-violating contributions to
$\delta Q_W^p/Q_W^p$ and $\delta Q_W^e/Q_W^e$, along with proposed experimental uncertainties.
\begin{table}
\begin{tabular}{|c|cccc|c|}
\hline
Quantity & $\Delta_{11k}^{\prime}(\tilde{d}_R^k)$
& $\Delta_{1k1}^{\prime}(\tilde{q}_L^k)$
& $\Delta_{12k}(\tilde{e}_R^k)$
& $\Delta_{21k}^{\prime}(\tilde{d}_R^k)$
& Value
\\
\hline\hline
$\delta |V_{ud}|^2/|V_{ud}|^2$
&2&0&-2&0&$-0.0029\pm 0.0014$
\\
$\delta Q_W^{\rm Cs}/Q_W^{\rm Cs}$
&-4.82&5.41&0.05&0&$-0.0040\pm 0.0066$
\\
$\delta R_{e/\mu}$
&2&0&0&-2&$-0.0042 \pm 0.0033$
\\
$\delta G_\mu/G_\mu$
&0&0&1&0&$0.00025\pm 0.001875$
\\
$\delta Q_W^p/Q_W^p$
&55.9&-27.9&-18.7&0&$\pm 0.040$
\\
$\delta Q_W^e/Q_W^e$
&0&0&-29.8&0&$\pm0.089$
\\
\hline
\end{tabular}
\caption{$P_R$-violating contributions to $\delta |V_{ud}|^2/|V_{ud}|^2$,
$\delta Q_W^{\rm Cs}/Q_W^{\rm Cs}$,
$\delta R_{e/\mu}$, $\delta G_\mu/G_\mu$, $\delta Q_W^p/Q_W^p$,
and $\delta Q_W^e/Q_W^e$.
Columns give the coefficients of the various corrections
from $\Delta_{ijk}^{\prime}$
and $\Delta_{12k}$ to the different quantities.
The last column gives the experimentally  measured value of the corresponding
quantity (for $Q_W^{p,e}$, only the proposed experimental uncertainties are shown).}
\label{tab:rpv-constr}
\end{table}
%

%As evident from Table~\ref{tab:rpv-constr}, only $\delta R_{e/\mu}$ depends on $\Delta_{21k}^{\prime}(\tilde{d}_R^k)$.
%Since $\Delta_{11k}^{\prime}(\tilde{d}_R^k)$ and $\Delta_{21k}^{\prime}(\tilde{d}_R^k)$
%appear with opposite signs, $\Delta_{21k}^{\prime}(\tilde{d}_R^k)$ can always
%be adjusted so that the contribution to the total $\chi^2$ from $R_{e/\mu}$ is zero, effectively eliminating
%this observable from the list of constraints.
%Therefore, we essentially have a three-parameter fit to $\delta |V_{ud}|^2/|V_{ud}|^2$,
%$\delta Q_W^{\rm Cs}/Q_W^{\rm Cs}$, and $\delta G_\mu/G_\mu$.
The 95\% CL region allowed by this fit in the $\delta Q_W^p/Q_W^p$ vs. $\delta Q_W^e/Q_W^e$ plane is
shown by the closed curve in Fig.~\ref{fig:mssm-vs-parity}. Note that the sign requirements
$\Delta_{ijk}(\tilde f),~\Delta_{ijk}^\prime(\tilde f)\ge 0$
[see Eq.~(\ref{eq:deltas})] truncate the initially elliptical curve to the shape shown in the figure.
%In particular, it is the
%condition $\Delta_{12k}(\tilde e_R^k)\ge 0$ that leads to $\delta Q_W^e/Q_W^e\le 0$ in the $P_R$-violating
%case [see Eq.~(\ref{eq:rpv-weak})].
We observe that the prospective effects of
$P_R$ non-conservation are quite distinct from SUSY loops. The value of $\delta Q_W^e/Q_W^e$ is
never positive in contrast to the situation for SUSY loop effects, whereas $\delta Q_W^p/Q_W^p$
can have either sign. Note, however, that the area enclosed by the curve corresponding to
$\delta Q_W^p/Q_W^p\ge 0$ is larger than the area corresponding to
$\delta Q_W^p/Q_W^p<0$, implying that $\delta Q_W^p/Q_W^p$ is more likely to be positive.
In addition, the magnitude of the $P_R$-violating effects can be roughly twice as large as possible magnitude of SUSY
loop effects for both $Q_W^{e,p}$. Thus, a comparison of results for the two
parity-violating electron scattering experiments could help
determine whether this extension of the MSSM is to be favored over other new
physics scenarios (see also Ref.~\cite{Erl-MJRM-Kur02}).

\section{Conclusions}
\label{sec:conclusion}

A new generation of precise, PVES experiments are poised to probe a variety
of scenarios for
physics beyond the SM \cite{slac,qweak,Mar96,Erl-MJRM-Kur02}. The
sensitivity of these measurements
to new physics is enhanced because the SM values for the electron and
proton weak charges are suppressed
and because theoretical uncertainties in the SM predictions are
sufficiently small\cite{Mar96,Erl-MJRM-Kur02}.
Here, we have studied the ability of these measurements to shed new light
on supersymmetric extensions of the SM.
We have observed that in a $P_R$-conserving version of the MSSM, the
effects of SUSY loop corrections to the
electron and proton weak charges are highly correlated and have the same
relative sign (positive) compared to the SM prediction over nearly
all the available MSSM parameter space. This correlation arises because the
corrections are dominated by
the SUSY loop contributions to  $\sin^2\theta^{eff}_W(q^2)$ -- a result
that would not have been obvious
in the absence of an explicit calculation. Moreover, the appearance of this
correlation does not result from
the adoption of any model for SUSY-breaking mediation, as we have
undertaken a model-independent analysis in
this study. We also find that the impact of SUSY radiative corrections on
the cesium weak charge are quite
small, so that the present agreement of $Q_W^{\rm Cs}$ with the SM does not
rule out potentially observable
effects in PVES.

In contrast, the effects on $Q_W^e$ and $Q_W^p$ induced by new tree-level,
$P_R$ violating SUSY interactions
display a different behavior. Given the constraints from other precision
electroweak observables, such as
the Fermi constant, first row CKM unitarity, and $Q_W^{\rm Cs}$, one would expect $P_R$
violation to cause a decrease in the size of $Q_W^e$. On the other hand, the magnitude of $Q_W^p$
can change either way, with an increase being more likely. Moreover,
the size of the $P_R$ violating
corrections could be even larger than those induced by SUSY loops,
particularly in the case of $Q_W^p$.
Should measurements of the weak charges be consistent with this
signature of $P_R$ violation, they could have important
implications for the nature of cold dark matter (it would not be
supersymmetric) and the nature of
neutrinos (they would be Majorana fermions).

From either standpoint, should the PVES measurements deviate significantly
from the SM predictions, one
may be able to draw interesting conclusions about the character of SUSY.
But what if both measurements turn
out to be consistent with the SM? In this case, one would add further
constraints to the possibility of
$P_R$ violation, but only marginally constrain the MSSM parameter
space based on possible loop effects.
In the latter case, the impact on both $Q_W^p$ and $Q_W^e$ is dominated by
$\delta\sin^2{\theta}_W^{eff}(q^2)^{\rm SUSY}$. Although the projected, combined
statistics of the two measurements
would make them more sensitive to SUSY radiative corrections than either measurement would
be independently, additional precision would be advantageous. In this respect, a possible
future measurement of $Q_W^e$ with a factor of two better precision than
anticipated at SLAC would
significantly enhance the ability of PVES to shed new light on
SUSY\footnote{We thank D. Mack, P. Reimer, and
P. Souder for sharing with us the possibility of such a future measurement
at the Jefferson Lab.}.

\begin{acknowledgments}
We thank R. Carlini, J. Erler, R. D. McKeown, and M. Wise for useful discussions.
This work is supported in part under U.S. Department of
Energy contract \#DE-FG03-02ER41215 (A.K. and M.J.R.-M.),
\#DE-FG03-00ER41132 (M.J.R.-M.), and \#DE-FG03-92-ER-40701 (S.S.).
A.K. and M.J.R.-M. are supported by
the National Science Foundation under award PHY00-71856.
S.S. is supported by the John A. McCone Fellowship.
\end{acknowledgments}

%-------------------------------------------------------------------------

%%%%%%%%%%%%%%%%%%%%%%%%%%%%%%%%%%%%%%%%%%%%%%%%%%%%%
%%%%%%%%%%%%%%%%%%%%%%%%%%%%%%%%%%%%%%%%%%%%%%%%%%%%%

%figures

%%%%%%%%%%%%%%%%%%%%%%%%%%%%%%%%%%%%%%%%%%%%%%%%%%%%%

\appendix

\section{Counterterms for the Effective PVES Lagrangian}
\label{app:counterterms}

The \lq\lq bare" effective Lagrangian for the forward angle PVES scattering
has the form:
\begin{eqnarray}
{\cal L}_{ef}&=&-{G_\mu^0\over {2\sqrt 2}}Q_W^{f0}
A^{\mu 0}_e\times V_{\mu f}^0 ~,\nonumber \\
{G_\mu^0\over {\sqrt 2}}&=&{g_0^2\over {8(M^0_W)^2}}
={G_\mu+\delta {\hat G}_\mu\over{\sqrt 2}} ~,\nonumber \\
Q_W^{f0}&=&2I_3^f-4Q_f s_0^2
=2I_3^f-4Q_f ({\hat s}^2+\delta {\hat s}^2) ~,\nonumber \\
A^{\mu 0}_e &=& \left({\bar e}\gamma^\mu\gamma_5 e \right)_0
\equiv A^{\mu}_e\left(1+{\delta  {\hat A}_e\over Q_W^f}\right) ~,\nonumber \\
V_{\mu f}^0 &=& \left({\bar f}\gamma_\mu f\right)_0
\equiv V_{\mu f}\left(1+{\delta {\hat V}_{f}\over Q_W^f}\right)
~,\end{eqnarray}
where the bare quantities are indexed by \lq\lq 0". Unless otherwise
indicated, all
higher-order contributions include both the SM and the SUSY pieces. The
quantity $1/Q_W^f$
in the parentheses in the last two lines of the above equation is
explicitly factored
out to make the definitions in Eq.~(\ref{eq:vertices-counter}) below more
convenient.

The counterterm $\delta {\hat G}_\mu$
is entirely determined by the muon lifetime. It can be taken from Eq. (62)
in Ref.~\cite{pierce}:
\begin{equation}
\label{eq:Gmu-counter}
{\delta {\hat G}_\mu \over G_\mu}=-{{\hat \Pi}_{WW}(0)\over M_W^2}-{\hat
\delta}_{VB}^\mu
~,\end{equation}
where ${\hat \Pi}_{WW}(q^2)$ is the $W$ boson self-energy and ${\hat
\delta}_{VB}^\mu$
is the sum of the vertex and box corrections to the muon decay amplitude.

We use the following convention for the $Z$-fermion interaction:
\begin{eqnarray}
\label{eq:gs}
V_{Zf}&=&-{g\over 4c}{\bar f}\gamma_\mu\left( g_V^f+g_A^f \gamma_5 \right)
f Z^\mu ~,\nonumber \\
g_V^f&=&2I_3^f-4Q_f s^2 ~,\nonumber \\
g_A^f&=&-2I_3^f
~.\end{eqnarray}
In this convention, the counterterms for the vector and the axial vector
currents can be read
off from Ref.~\cite{pierce}
\footnote{
Note that Ref.~\cite{pierce} has the opposite sign convention for $g_A^f$.
}:
\begin{eqnarray}
\label{eq:vertices-counter}
\delta {\hat A}_e&=& -g_V^e \delta{\hat Z_A^e}+g_A^e \delta{\hat Z_V^e}
~,\nonumber \\
\delta {\hat V}_f&=& g_V^f  \delta{\hat Z_V^f}-g_A^f  \delta{\hat Z_A^f}
~,\nonumber \\
\delta{\hat Z_A^{e,f}}&=&{1\over 2}\left(\delta{\hat Z_L^{e,f}}
-\delta{\hat Z_R^{e,f}}\right) ~,\nonumber \\
\delta{\hat Z_V^{e,f}}&=&{1\over 2}\left(\delta{\hat Z_L^{e,f}}
+\delta{\hat Z_R^{e,f}}\right)
~,\end{eqnarray}
where $\delta{\hat Z_L^{e,f}}$ and $\delta{\hat Z_R^{e,f}}$ are the field
strength
renormalization constants for left- and right-handed fermions, respectively.
One can write the one-loop correction to the NC vertex as
\begin{equation}
-(ig/4c) \delta {\hat V}_\mu^f=-(ig/4c){\bar
f}\gamma_\mu\left({\hat G}_V^f+\gamma_5{\hat G}_A^f\right)f
~,\end{equation}
where only the contributions that are not suppressed by powers of either
the momentum transfer ($\sqrt{|q^2|}/M_{\rm SUSY}$)
or the fermion mass ($m_f/M_{\rm SUSY}$) are shown.
The quantities ${\hat G}_A^e$ and ${\hat G}_V^f$ represent rescaling of the vertices
by the one-loop radiative corrections.
They must be combined with the appropriate
counterterms from Eq.~(\ref{eq:vertices-counter}) to obtain the
renormalized corrections:
\begin{eqnarray}
\label{eq:vert-renorm}
{\hat V}_A^e&=&{\hat G}_A^e+\delta {\hat A}_e={\hat G}_A^e
-g_V^e \delta{\hat Z_A^e}+g_A^e \delta{\hat Z_V^e} ~,\nonumber \\
{\hat V}_V^f&=&{\hat G}_V^f+\delta {\hat V}_f={\hat G}_V^f
+g_V^f \delta{\hat Z_V^f}-g_A^f \delta{\hat Z_A^f}
~.\end{eqnarray}

\section{Decoupling of Gluinos from the Weak Charge of Quarks}
\label{sec:glue-dec}

It is sufficient to demonstrate the decoupling for one of the quark flavors
({\it e.g.} the up quark)
since for other flavors the
proof is identical. Consider the renormalized vector neutral current vertex
for the up
quark [see Eq.~(\ref{eq:vert-renorm})]:
\begin{eqnarray}
\label{eq:vert-ren}
{\hat V}_V^u&=&{\hat G}_V^u+g_V^u \delta{\hat Z_V^u}-g_A^u \delta{\hat Z_A^u} ~,\nonumber \\
\delta{\hat Z_V^u}&=&{\delta{\hat Z_L^u}+\delta{\hat Z_R^u}\over 2}
~,\nonumber \\
\delta{\hat Z_A^u}&=&{\delta{\hat Z_L^u}-\delta{\hat Z_R^u}\over 2}
~.\end{eqnarray}
The gluino contributions to ${\hat G}_V^u$ are given by the graph shown in Fig.
\ref{fig:nc-uquark-vertex}(b),
with the neutralino replaced by the gluino. By using Eq.
(\ref{eq:nc-uquark-vertb-expr}) below with the
appropriate coupling constants it is straightforward to show that:
\begin{eqnarray}
\label{eq:nc-ugluino}
{\hat G}_V^u({\rm Gluino})&=&-{4\over 3}{{\alpha_S}\over{4 \pi }}{\sum_{i,j}}
\left({\sum_{I}}Z_U^{*Ii}Z_U^{Ij}-{4\over 3}{\hat s}^2\delta_{ij}\right) \nonumber \\
&\times& \left [ g_{GL}^{*uj}g_{GL}^{ui}+g_{GR}^{*uj}g_{GR}^{ui} \right]
V_2(M_{\tilde g},m_{\tilde U_i},m_{\tilde U_j})
~,\end{eqnarray}
where $\alpha_S$ is the strong coupling constant, $M_{\tilde g}$ is the
gluino mass, $V_2(M,m_1,m_2)$ is
defined in Eq.~(\ref{eq:vert-raw}) below and (see Ref.~\cite{kur-rm-su-nutev})
\begin{eqnarray}
\label{eq:glue-copul}
g_{GL}^{ui}&=&-\sqrt{2}Z^{*1i}_U ~,\nonumber \\
g_{GR}^{ui}&=&\sqrt{2}Z^{*4i}_U
~.\end{eqnarray}
In this work, no flavor mixing in the squark sector is allowed. Therefore,
$Z^{1i}_U,Z^{4i}_U\ne 0$ only if
$i=1,4$. Since $Z_U$ is unitary we find:
\begin{equation}
\label{eq:closure}
g_{GL}^{*uj}g_{GL}^{ui}+g_{GR}^{*uj}g_{GR}^{ui}=2\left(Z^{1j}_UZ^{*1i}_U+Z^{4j}_UZ^{*4i}_U\right)=2\delta_{ij}
\end{equation}
for $i,j=1,4$. Finally:
\begin{eqnarray}
\label{eq:GVglue-final}
{\hat G}_V^u({\rm Gluino})&=&-{4\over 3}{{\alpha_S}\over{2 \pi }}{\sum_{i=1,4}}
\left(|Z_U^{1i}|^2-{4\over 3}s^2\right) V_2(M_{\tilde g},m_{\tilde U_i},m_{\tilde U_i})
\nonumber \\
&=&-{4\over 3}{{\alpha_S}\over{2 \pi }}
\Biggl\{
{g_V^u\over 2}
\left[V_2(M_{\tilde g},m_{\tilde U_1},m_{\tilde U_1})+V_2(M_{\tilde
g},m_{\tilde U_4},m_{\tilde U_4})\right]
\nonumber \\
&-&{g_A^u\over 2}\left(1-2|Z_U^{14}|^2 \right)
\left[V_2(M_{\tilde g},m_{\tilde U_1},m_{\tilde U_1})-V_2(M_{\tilde
g},m_{\tilde U_4},m_{\tilde U_4})\right]
\Biggr\}
~,\end{eqnarray}
where the closure property Eq.~(\ref{eq:closure}) was used together with
\begin{eqnarray}
g_V^u&=&2(I_3^u-2Q^u {\hat s}^2)=1-{{8\over 3} {\hat s}^2} ~,\nonumber \\
g_A^u&=&-2I_3^u=-1
~.\end{eqnarray}
On the other hand, the gluino-induced wave function renormalization
constants of the up quark have the form:
\begin{eqnarray}
\delta{\hat Z_V^u}({\rm Gluino})&=&{4\over 3}{{\alpha_S}\over{8 \pi }}
\sum_{i}
\left ( |g_{GL}^{ui}|^2+|g_{GR}^{ui}|^2 \right )
F_1(m_{\tilde U_i},M_{\tilde g},0) ~,\nonumber \\
\delta{\hat Z_A^u}({\rm Gluino})&=&{4\over 3}{{\alpha_S}\over{8 \pi }}
\sum_{i}
\left ( |g_{GL}^{ui}|^2-|g_{GR}^{ui}|^2 \right )
F_1(m_{\tilde U_i},M_{\tilde g},0)
~,\end{eqnarray}
where $F_1(m_1,m_2,m_3)$ is given by
\begin{equation}
\label{eq:F-function}
F_1(m_1,m_2,m_3)={\int_0^1}x \ln\left\{\left[x m_1^2+(1-x)m_2^2-x(1-x)m_3^2\right]/\mu^2\right\}
~.\end{equation}
Note that according to Eq.~(\ref{eq:vert-raw}) below, $F_1(m_1,m_2,0)\equiv
V_2(m_2,m_1,m_1)$. Using
Eqs. (\ref{eq:glue-copul}) and (\ref{eq:closure}) we find
\begin{eqnarray}
\label{eq:Zglue-final}
\delta{\hat Z_V^u}({\rm Gluino})&=&{4\over 3}{{\alpha_S}\over{4 \pi }}
\left[V_2(M_{\tilde g},m_{\tilde U_1},m_{\tilde U_1})+V_2(M_{\tilde
g},m_{\tilde U_4},m_{\tilde U_4})\right] ~,\nonumber \\
\delta{\hat Z_A^u}({\rm Gluino})&=&{4\over 3}{{\alpha_S}\over{4 \pi }}
\left (1-2|Z_U^{14}|^2 \right )
\left[V_2(M_{\tilde g},m_{\tilde U_1},m_{\tilde U_1})-V_2(M_{\tilde
g},m_{\tilde U_4},m_{\tilde U_4})\right]
~.\end{eqnarray}
After substitution of Eqs. (\ref{eq:GVglue-final}) and
(\ref{eq:Zglue-final}) into Eq.~(\ref{eq:vert-ren})
the gluino corrections to the vector neutral current vertex of the up quark
cancel
exactly. Therefore, gluino loops do not renormalize the weak charge of the
up quark.

\section{Complete Expressions For Feynman Diagrams}
\label{app:expressions}

In this appendix we list analytical expressions for all SUSY one-loop
vertex and box
Feynman diagrams that contribute to PV electron scattering. The complete
expressions for
remaining diagrams (see Section \ref{sec:general}) are given in the appendices of Ref.
\cite{kur-rm-su-nutev}. We use the capitalized letters $I$ and $J$ to denote 
the family index for quarks and leptons ($I,J=1,\cdots,3$), 
small letters $i$ and $j$ to denote the index for squarks and sleptons 
($i,j=1,\cdots,6$ except for sneutrino, when $i,j=1,\cdots,3$), and small 
letters $p$ and $n$ to denote the index for the neutralinos ($p,n=1,\cdots,4$)
and charginos ($p,n=1,2$).

%%%%%%%%%%%%%%%%%%%%%%%%%%%%%%%%%%%%%%%%%%%%%%%%%%%%%

\subsection{Vertex Corrections}

The Feynman diagrams are shown in Figs. \ref{fig:nc-electron-vertex} and
\ref{fig:nc-uquark-vertex}. Let us start with the corrections to the
$e-e-Z$ vertex. The loop integral functions $V_1(m_1,m_2,m_3)$ and
$V_2(m_1,m_2,m_3)$ are defined as
\begin{eqnarray}
\label{eq:vert-raw}
V_1(M,m_1,m_2)&=&\int_0^1dx\int_0^1dy{y\over D_3(M,m_1,m_2)}~,\nonumber \\
V_2(M,m_1,m_2)&=&\int_0^1dx\int_0^1dy{y\ln \left[D_3(M,m_1,m_2)/\mu^2\right]}~,\nonumber \\
D_3(M,m_1,m_2)&=&(1-y)M^2+y[(1-x)m_1^2+xm_2^2]
~,\end{eqnarray}
where $\mu$ is the renormalization scale.
Explicitly:
\begin{eqnarray}
\label{eq:vert-functions}
V_1(M,m_1,m_2)&=&{{m_1^2\ln{m_1^2\over M^2}}
\over{(M^2-m_1^2)(m_2^2-m_1^2)}}
+{{m_2^2\ln{m_2^2\over M^2}}
\over{(M^2-m_2^2)(m_1^2-m_2^2)}}
~,\nonumber \\
V_2(M,m_1,m_2)&=&{1\over 4}\Biggl[2\ln{M^2}-3
+{2m_1^4\over{(M^2-m_1^2)(m_2^2-m_1^2)}}\ln{m_1^2\over M^2}
\nonumber \\
&&+{2m_2^4\over{(M^2-m_2^2)(m_1^2-m_2^2)}}\ln{m_2^2\over M^2}
-2 \ln \mu^2\Biggr]
~.\end{eqnarray}
We have [${\hat P}_L=(1-\gamma_5)/2$, ${\hat P}_R=(1+\gamma_5)/2$]:
\begin{eqnarray}
\label{eq:nc-electron-verta-expr}
\delta {\hat V}_\mu^{e(a)}&=&-{{\alpha}\over{2 \pi}}{\sum_{i,j,p}}
\left({\sum_{I}}Z_\nu^{*Ii}Z_\nu^{Ij}-2Q_\nu
{\hat s}^2\delta_{ij}\right)
V_2(m_{\chi^+_p},m_{{\stilde \nu}_i},m_{{\stilde \nu}_j}) \nonumber \\
&\times&{\bar e} \gamma_\mu \left [
g_{L}^{*ejp}g_{L}^{eip}{\hat P}_L+g_{R}^{*ejp}g_{R}^{eip}{\hat P}_R
\right ] e
~.\end{eqnarray}
Note that $Q_\nu=0$ and $Z_\nu^{ij}$ is a unitary $3\times 3$
matrix. Therefore, one identically
has ${\sum_{I}}Z_\nu^{*Ii}Z_\nu^{Ij}-2Q_\nu {\hat s}^2\delta_{ij}=\delta_{ij}$.
The explicit form is kept
so that the down quark neutral current vertex
may be easily obtained by the replacement $e\rightarrow d$ (with $Z_L\to Z_D$)
and $\nu\to u$.
\begin{eqnarray}
\label{eq:nc-electron-vertb-expr}
\delta {\hat V}_\mu^{e(b)}&=&{{\alpha}\over{2 \pi}}{\sum_{i,j,p}}
\left({\sum_{I}}Z_L^{*Ii}Z_L^{Ij}+2Q_e{\hat s}^2\delta_{ij}\right)
V_2(m_{\chi^0_p},m_{\tilde L_i},m_{\tilde L_j}) \nonumber \\
&\times& {\bar e} \gamma_\mu \left [
g_{0L}^{*ejp}g_{0L}^{eip}{\hat P}_L+g_{0R}^{*ejp}g_{0R}^{eip}{\hat P}_R
\right ] e~,\\
\delta {\hat V}_\mu^{e(c)}&=&{{\alpha}\over{2 \pi}}{\sum_{i,p,n}}
{\bar e} \gamma_\mu
\Biggl\{\nonumber \\
&+&\left[
O_{pn}^{R\prime}g_{L}^{*ein}g_{L}^{eip}{\hat P}_L
+O_{pn}^{L\prime}g_{R}^{*ein}g_{R}^{eip}{\hat P}_R
\right] 2m_{\chi^+_p}m_{\chi^+_n}
V_1(m_{{\stilde \nu}_i},m_{\chi^+_p},m_{\chi^+_n})\nonumber \\
&-&\left[
O_{pn}^{L\prime}g_{L}^{*ein}g_{L}^{eip}{\hat P}_L
+O_{pn}^{R\prime}g_{R}^{*ein}g_{R}^{eip}{\hat P}_R
\right] \left[
1+2 V_2(m_{{\stilde \nu}_i},m_{\chi^+_p},m_{\chi^+_n})
\right]\Biggr\}e~,\\
\delta {\hat V}_\mu^{e(d)}&=&-{{\alpha}\over{2 \pi}}{\sum_{p,n,i}}
{\bar e} \gamma_\mu \Biggl\{\nonumber \\
&+& \left[
O_{np}^{L\prime\prime}g_{0L}^{*ein}g_{0L}^{eip}{\hat P}_L
+O_{np}^{R\prime\prime}g_{0R}^{*ein}g_{0R}^{eip}{\hat P}_R
\right] 2m_{\chi^0_p}m_{\chi^0_n}
V_1(m_{\tilde L_i},m_{\chi^0_p},m_{\chi^0_n})\nonumber \\
&-&\left[
O_{np}^{R\prime\prime}g_{0L}^{*ein}g_{0L}^{eip}{\hat P}_L
+O_{np}^{L\prime\prime}g_{0R}^{*ein}g_{0R}^{eip}{\hat P}_R
\right] \left[
1+2 V_2(m_{\tilde L_i},m_{\chi^0_p},m_{\chi^0_n})
\right]\Biggr\}e
~.\end{eqnarray}

The vector and axial vector pieces can be readily read off from the above
formulae. The radiative corrections to the up quark neutral current vertex are
as follows (see Fig.~\ref{fig:nc-uquark-vertex}):
\begin{eqnarray}
\label{eq:nc-uquark-verta-expr}
\delta {\hat V}_\mu^{u(a)}&=&{{\alpha}\over{2 \pi}}{\sum_{i,j,p}}
\left({\sum_{I}}Z_D^{*Ii}Z_D^{Ij}-{2\over 3}{\hat s}^2\delta_{ij}\right)
V_2(m_{\chi^+_p},m_{\tilde D_i},m_{\tilde D_j}) \nonumber \\
&\times& {\bar u} \gamma_\mu \left [
g_{L}^{*ujp}g_{L}^{uip}{\hat P}_L+g_{R}^{*ujp}g_{R}^{uip}{\hat P}_R
\right ] u~,\\
\label{eq:nc-uquark-vertb-expr}
\delta {\hat V}_\mu^{u(b)}&=&-{{\alpha}\over{2 \pi}}{\sum_{i,j,p}}
\left({\sum_{I}}Z_U^{*Ii}Z_U^{Ij}-{4\over 3}{\hat s}^2\delta_{ij}\right)
V_2(m_{\chi^0_p},m_{\tilde U_i},m_{\tilde U_j}) \nonumber \\
&\times& {\bar u} \gamma_\mu \left [
g_{0L}^{*ujp}g_{0L}^{uip}{\hat P}_L+g_{0R}^{*ujp}g_{0R}^{uip}{\hat P}_R
\right ] u~,\\
\delta {\hat V}_\mu^{u(c)}&=&-{{\alpha}\over{2 \pi}}{\sum_{i,p,n}}
{\bar u} \gamma_\mu
\Biggl\{\nonumber \\
&+& \left[
O_{np}^{L\prime}g_{L}^{*uin}g_{L}^{uip}{\hat P}_L
+O_{np}^{R\prime}g_{R}^{*uin}g_{R}^{uip}{\hat P}_R
\right] 2m_{\chi^+_p}m_{\chi^+_n}
V_1(m_{\tilde D_i},m_{\chi^+_p},m_{\chi^+_n})\nonumber \\
&-&\left[
O_{np}^{R\prime}g_{L}^{*uin}g_{L}^{uip}{\hat P}_L
+O_{np}^{L\prime}g_{R}^{*uin}g_{R}^{uip}{\hat P}_R
\right] \left[
1+2 V_2(m_{\tilde D_i},m_{\chi^+_p},m_{\chi^+_n})
\right]\Biggr\}u~,\\
\delta {\hat V}_\mu^{u(d)}&=&-{{\alpha}\over{2 \pi}}{\sum_{p,n,i}}
{\bar u} \gamma_\mu \Biggl\{\nonumber \\
&+& \left[
O_{np}^{L\prime\prime}g_{0L}^{*uin}g_{0L}^{uip}{\hat P}_L
+O_{np}^{R\prime\prime}g_{0R}^{*uin}g_{0R}^{uip}{\hat P}_R
\right] 2m_{\chi^0_p}m_{\chi^0_n}
V_1(m_{\tilde U_i},m_{\chi^0_p},m_{\chi^0_n})\nonumber \\
&-&\left[
O_{np}^{R\prime\prime}g_{0L}^{*uin}g_{0L}^{uip}{\hat P}_L
+O_{np}^{L\prime\prime}g_{0R}^{*uin}g_{0R}^{uip}{\hat P}_R
\right] \left[
1+2 V_2(m_{\tilde U_i},m_{\chi^0_p},m_{\chi^0_n})
\right]\Biggr\}u
~.\end{eqnarray}

\subsection{Anapole Moment Corrections}

Using formulae in the appendix of Ref.~\cite{kur-rm-su-nutev} we find for
$F_A^e(0)$
of the electron:
\begin{eqnarray}
F_A^e(0)&=&F_A^{(a)}(0)+F_A^{(b)}(0) ~,\nonumber \\
F_A^{(a)}(0)&=&-{\alpha M_Z^2\over{48\pi}}\sum_{i,p}|g_L^{eip}|^2\int_0^1
{x^2(x-3)dx\over{(1-x)m_{{\stilde \nu}_i}^2+x m_{\chi^+_p}^2}}~, \\
F_A^{(b)}(0)&=&{\alpha M_Z^2\over{48\pi}}\sum_{i,p}
\left(|g_{0L}^{eip}|^2-|g_{0R}^{eip}|^2\right)
\int_0^1 {x^3dx\over{(1-x)m_{\chi^0_p}^2+x m_{\tilde L_i}^2}}
~.\end{eqnarray}

\subsection{Box Graphs}

Let us introduce the following notation:
\begin{eqnarray}
L_{fi}&=&{\bar f}(p_i^\prime)\gamma_\mu(1-\gamma_5)f(p_i) ~,\nonumber \\
R_{fi}&=&{\bar f}(p_i^\prime)\gamma_\mu(1+\gamma_5)f(p_i)
~.\end{eqnarray}
As explicitly shown below, each box graph has the following structure:
\begin{eqnarray}
M_{Box}^{ef}&=&i{G_\mu\over {\sqrt 2}}\Biggl(
A^{ef} L_{f2}\times L_{e1}
+B^{ef} R_{f2}\times R_{e1} \nonumber \\
&+&C^{ef} R_{f2}\times L_{e1}
+D^{ef} L_{f2}\times R_{e1}\Biggr)
~.\end{eqnarray}
To study the effects of the parity-violating electron scattering we need to
pick
only the term that has axial vector current on the projectile side $e1$ and
the vector
current on the target side $f2$. Therefore, it is easily seen that the box
diagram
contribution to $C_{1f}$ is [see Eq.~(\ref{eq:nc-lambda})]:
\begin{equation}
\label{eq:del-box}
{\hat\delta}_{Box}^{ef}=-2(-A^{ef}+B^{ef}-C^{ef}+D^{ef})
\end{equation}
in the above notation. The explicit expressions for the $ee$ box graphs
in Fig.~\ref{fig:nc-electron-boxes} are given below.
\begin{eqnarray}
\label{eq:nc-ee-boxa-expr}
\delta M_{Box}^{ee(a)}&=&i{G_\mu\over {\sqrt 2}}{\alpha M_W^2 {\hat s}^2 \over {4 \pi}}
{\sum_{n,p,i,j}} \nonumber \\
&&\Biggl\{
\Biggl[
g_{0L}^{*ejn}g_{0L}^{ejp}g_{0L}^{*ein}g_{0L}^{eip} L_{e2}\times L_{e1}
+g_{0R}^{*ejn}g_{0R}^{ejp}g_{0R}^{*ein}g_{0R}^{eip} R_{e2}\times R_{e1}
\Biggr]\nonumber \\
&\times& m_{\chi^0_p} m_{\chi^0_n}
B_2(m_{\tilde L_i},m_{\tilde L_j},m_{\chi^0_p},m_{\chi^0_n})
\nonumber \\
&+& \Biggl[
g_{0R}^{*ejn}g_{0R}^{ejp}g_{0L}^{*ein}g_{0L}^{eip} R_{e2}\times L_{e1}
+g_{0L}^{*ejn}g_{0L}^{ejp}g_{0R}^{*ein}g_{0R}^{eip} L_{e2}\times R_{e1}
\Biggr]\nonumber \\
&\times& B_1(m_{\tilde L_i},m_{\tilde L_j},m_{\chi^0_p},m_{\chi^0_n})\Biggr\}~,\\
\delta M_{Box}^{ee(b)}&=&-i{G_\mu\over {\sqrt 2}}{\alpha M_W^2 {\hat s}^2 \over {4 \pi}}
{\sum_{n,p,i,j}} \nonumber \\
&&\Biggl\{
\Biggl[
g_{0L}^{*ejn}g_{0L}^{ejp}g_{0L}^{ein}g_{0L}^{*eip} L_{e2}\times L_{e1}
+g_{0R}^{*ejn}g_{0R}^{ejp}g_{0R}^{ein}g_{0R}^{*eip} R_{e2}\times R_{e1}
\Biggr]\nonumber \\
&\times& B_1(m_{\tilde L_i},m_{\tilde L_j},m_{\chi^0_p},m_{\chi^0_n})
\nonumber \\
&+& \Biggl[
g_{0R}^{*ejn}g_{0R}^{ejp}g_{0L}^{ein}g_{0L}^{*eip} R_{e2}\times L_{e1}
+g_{0L}^{*ejn}g_{0L}^{ejp}g_{0R}^{ein}g_{0R}^{*eip} L_{e2}\times R_{e1}
\Biggr]\nonumber \\
&\times& m_{\chi^0_p} m_{\chi^0_n}
B_2(m_{\tilde L_i},m_{\tilde L_j},m_{\chi^0_p},m_{\chi^0_n})\Biggr\}~,\\
\delta M_{Box}^{ee(c)}&=&-i{G_\mu\over {\sqrt 2}}{\alpha M_W^2 {\hat s}^2\over {4 \pi }}
{\sum_{n,p,i,j}}
g_{L}^{*eip}g_{L}^{ein}g_{L}^{ejp}g_{L}^{*ejn} L_{e2}\times L_{e1}
\nonumber \\
&\times& B_1(m_{{\stilde \nu}_i},m_{{\stilde \nu}_j},m_{\chi^+_p},m_{\chi^+_n})
~.\end{eqnarray}
In all above formulae the following functions are used:
\begin{eqnarray}
\label{eq:cc-quark-boxd-expr}
B_1(M_1,M_2,m_1,m_2)&=&{\int_0^1dx}{\int_0^1dy}{\int_0^1dz}{{z(1-z)}
\over D_4(M_1,M_2,m_1,m_2)} ~,\nonumber \\
B_2(M_1,M_2,m_1,m_2)&=&{\int_0^1dx}{\int_0^1dy}{\int_0^1dz}{{z(1-z)}
\over D_4^2(M_1,M_2,m_1,m_2)} ~,\nonumber \\
D_4(M_1,M_2,m_1,m_2)&=&z[(1-x)M_1^2+xM_2^2]
+(1-z)[(ym_1^2+(1-y)m_2^2]
~.\end{eqnarray}
Explicitly:
\begin{eqnarray}
\label{eq:box-integrals}
&&B_1(M_1,M_2,m_1,m_2)={m_1^4\ln{m_1^2\over M_2^2}
\over{2(m_1^2-M_1^2)(m_1^2-m_2^2)(m_1^2-M_2^2)}} \nonumber \\
&&+{m_2^4\ln{m_2^2\over M_2^2}
\over{2(m_2^2-m_1^2)(m_2^2-M_1^2)(m_2^2-M_2^2)}} +
{M_1^4\ln{M_1^2\over M_2^2}
\over{2(M_1^2-m_1^2)(M_1^2-m_2^2)(M_1^2-M_2^2)}}~, \nonumber \\
&&B_2(M_1,M_2,m_1,m_2)={m_1^2\ln{M_2^2\over m_1^2}
\over{(m_1^2-M_1^2)(m_1^2-m_2^2)(m_1^2-M_2^2)}} \nonumber \\
&&+{m_2^2\ln{M_2^2\over m_2^2}
\over{(m_2^2-m_1^2)(m_2^2-M_1^2)(m_2^2-M_2^2)}} +
{M_1^2\ln{M_2^2\over M_1^2}
\over{(M_1^2-m_1^2)(M_1^2-m_2^2)(M_1^2-M_2^2)}}
~.\end{eqnarray}

The box graphs for the electron-down quark scattering are easily obtained from
the above expressions by replacing the target electron $e2$ with the down
quark.
Also, all quantities that have the running index $j$ re to be replaced
with the corresponding quantities for the first generation {\it down} squarks:
$g_{0L}^{ejn}\to g_{0L}^{djn}$, {\it etc}.

The box graphs for the electron-up quark scattering are shown in Fig.
\ref{fig:nc-uquark-boxes}. The graphs (a) and (b) are easily obtained from
the corresponding graphs for the $ee$ scattering by replacing all
quantities that have the running index $i$ with the corresponding quantities
for the first {\it up} generation squarks:
$g_{0L}^{ein}\rightarrow g_{0L}^{uin}$, {\it etc}. The result for the
last graph (c) is:
\begin{eqnarray}
\label{eq:nc-eu-boxc-expr}
\delta M_{Box}^{eu(c)}&=&i{G_\mu\over {\sqrt 2}}{\alpha M_W^2 {\hat s}^2 \over {4 \pi }}
{\sum_{n,p,i,j}}
g_{L}^{*ujn}g_{L}^{ujp}g_{L}^{*ein}g_{L}^{eip} L_{u2}\times L_{e1}
\nonumber \\
&\times& m_{\chi^+_p} m_{\chi^+_n}
B_2(m_{{\stilde \nu}_i},m_{\tilde D_j},m_{\chi^+_p},m_{\chi^+_n})
~.\end{eqnarray}


\begin{thebibliography}{99}

\bibitem{SLAC}C.Y. Prescott {\em et al.}, Phys. Lett. B {\bf 77}, 347 (1978);
{\em ibid} {\bf 84}, 524 (1979).

\bibitem{Ben99} S.C. Bennett and C.E. Wieman, Phys. Rev. Lett. {\bf 82},
2484 (1999); C.S. Wood {\em et al.}, Science {\bf 275}, 1759 (1997).

\bibitem{NuTeV} G.P. Zeller {\em et al}, NuTev Collaboration, Phys. Rev.
Lett. {\bf 88}:091802
(2002).

\bibitem{slac} SLAC Experiment E-158, E. W. Hughes, K. Kumar, P.A. Souder,
spokespersons.

\bibitem{qweak} JLab experiment E-02-020, R. Carlini, J.M. Finn, S. Kowalski,
and S. Page, spokespersons.

\bibitem{Erl-MJRM-Kur02} J. Erler, M.J. Ramsey-Musolf, and A. Kurylov,
[hep-ph/0302149] (2003).

\bibitem{kamion-cosm} G. Jungman, M. Kamionkowski, and K. Griest, Phys.
Rep. {\bf 267}, 195 (1996).

\bibitem{haber-kane} H.E. Haber, G.L Kane, Phys. Rep. {\bf 117}, 74 (1985).

\bibitem{kurylov02} A. Kurylov and M.J. Ramsey-Musolf, Phys. Rev.
Lett. {\bf 88}, 071804 (2002).


\bibitem{Mar96}A. Czarnecki and W. Marciano, Phys. Rev. D {\bf 53}, 1066
(1996).


\bibitem{kur-rm-su-nutev} A. Kurylov, M. J. Ramsey-Musolf, and S. Su,
hep-ph/0301208.

\bibitem{PDG00} K. Hagiwara {\it et. al.}, Review of Particle Physics,
Phys. Rev. D {\bf 66}, 10001 (2002).

\bibitem{rosiek} J. Rosiek, Phys. Rev. D {\bf 41} (1990) 3464. Erratum
hep-ph/9511250.

\bibitem{Kane} See {\em e.g.} Perspectives on Supersymmetry, G.L. Kane, Ed.,
World Scientific, Singapore, 1998.

\bibitem{erler} J. Erler and D. M. Pierce, Nucl. Phys. B {\bf 526}, 53 (1998).

\bibitem{siegel}W. Siegel, Phys. Lett. B {\bf 84}, 193 (1979).

\bibitem{pierce} Damien M. Pierce, hep-ph/9805497.

\bibitem{sirlin-nc} W. J. Marciano and A. Sirlin, Phys. Rev. D {\bf 22},
2695 (1980)

\bibitem{stu-degrassi} G. Degrassi, B. A. Kniehl, and A. Sirlin,
Phys. Rev. D {\bf 48}, 3963 (1993).

\bibitem{haber93} H.E. Haber, hep-ph/9306207; S. Heinemeyer and G. Weiglein,
hep-ph/0102317.

\bibitem{rm-anapole} M. J. Musolf, B. R. Holstein, Phys. Rev. D {\bf 43},
2956 (1991).

\bibitem{MRM00}M.J. Ramsey-Musolf, Phys. Rev. D {\bf 62}:056009 (2000).

\bibitem{bnl-muon} G. W. Bennett {\it et. al.} (Muon $g-2$ Collaboration),
Phys. Rev. Lett. {\bf 89}, 101804 (2002).

\bibitem{MRM99}M.J. Ramsey-Musolf, Phys. Rev. C {\bf 60}:015501 (1999).

\bibitem{sushov02} A. I .Milstein, O.P. Sushkov, and I.S. Terekhov,
hep-ph/0212072.

\bibitem{towner-super} I. S. Towner and J. C. Hardy, {\em Proceedings of the
Fifth International WEIN Symposium: Physics Beyond the Standard Model}, P.
Herczeg, {\em et
al.}, eds. (World Scientific, 1999) p. 338.

\bibitem{pil2} G. Czapek {\em et al.}, Phys. Rev. Lett. {\bf 70}, 17
(1993); D. I. Britton, {\em et al.}, Phys. Rev. Lett. {\bf 68}, 3000 (1992).

\bibitem{marciano99} W. J. Marciano, Phys. Rev. {\bf D 60}:093006 (1999).

%\bibitem{barger89} V. Barger, G. F. Guidice, and T. Han, Phys. Rev. D {\bf
%40}, 2987 (1989).

\end{thebibliography}
\end{document}